# On two quantum approaches to adaptive mutations in bacteria.

**Vasily Ogryzko***
**INSERM, CNRS UMR 8126, Universite Paris Sud XI**
**Institut Gustave Roussy, Villejuif, France**

vogryzko@gmail.com

http://sites.google.com/site/vasilyogryzko/

ABSTRACT

The phenomenon of adaptive mutations has been attracting attention of biologists for several decades as challenging the basic premise of the Central Dogma of Molecular Biology. Two approaches, based on the quantum theoretical principles (*QMAMs* - Quantum Models of Adaptive Mutations) have been proposed in order to explain this phenomenon. In the present work, they are termed *Q-cell* and *Q-genome* approaches and are compared using 'fluctuation trapping' mechanism as a general framework. Notions of *R-error* and *D-error* are introduced, and it is argued that the 'fluctuation trapping model' can be considered as a QMAM only if it employs a correlation between the *R-* and *D-errors*. It is shown that the model of McFadden & Al-Khalili (1999) cannot qualify as a *QMAM*, as it corresponds to the *'D-error* only' model. Further, the paper compares how the *Q-cell* and *Q-genome* approaches can justify the *R-D-error* correlation, focusing on the advantages of the *Q-cell* approach. The positive role of environmentally induced decoherence (EID) on both steps of the adaptation process in the framework of the *Q-cell* approach is emphasized. A starving bacterial cell is proposed to be in an einselected state. The intracellular dynamics in this state has a unitary character and is proposed to be interpreted as 'exponential growth in imaginary time', analogously to the commonly considered 'diffusion' interpretation of the Schroedinger equation. Addition of a substrate leads to Wick rotation and a switch from 'imaginary time' reproduction to a 'real time' reproduction regime. Due to the variations at the genomic level (such as base tautomery), the starving cell has to be represented as a superposition of different components, all 'reproducing in imaginary time'. Any addition of a selective substrate, allowing only one of these components to amplify, will cause Wick rotation and amplification of this component, thus justifying the occurrence of the *R-D-error* correlation. Further ramifications of the proposed ideas for evolutionary theory are discussed.

# 1. Introduction

The exorcism of teleology from the natural sciences is widely considered to be the main legacy of Darwinism (Dennet, 1996). According to the Darwinian paradigm, the adaptation of life to its surroundings does not involve any 'foresight' and can be explained as the result of a random search through a succession of heritable variations and selection. The essential claim of Darwinism that an adaptive value of a heritable variation cannot be directly anticipated by an organism and can be proven only via selection on the populational level is supported by the Central Dogma of molecular biology (Crick, 1970). This widely accepted framework for understanding the mechanisms of gene expression forbids the transfer of sequential information from proteins to nucleic acids, erecting a border between genotype and phenotype and thus separating heritable variations from selection.

The notion that an individual organism cannot adapt directly to its environment by changing its genome has been challenged in the last two decades with the discovery that the emergence of some mutations in microorganisms depends on their phenotypic consequences, i.e., they appear mostly when they are needed for the cell to grow (Cairns et al., 1988; Foster, 2000; Hall, 1991; Roth et al., 2006). This 'phenomenon of adaptive mutations' is more in accordance with the views of Lamarck (Lamarck, 1809), who believed that individual organisms have enough plasticity to contribute directly into the evolutionary process.

From the physical point of view, life can be considered as a particular case of condensed matter (Anderson & Stein, 1987). Bearing with the fact that quantum mechanics (quantum field theory, in particular) is indispensable for understanding the physics of condensed matter, many authors, most notably Schroedinger (Schroedinger, 1944), have suggested that quantum principles have to play a role in the inner workings of life. An additional reason to believe that quantum theory will be required, is the ongoing progress of 'omics-' and 'nano-' technologies in biological sciences, which will eventually lead to a recognition of the limits to how much can be observed concerning an individual biological object (e.g., a single cell) (Ogryzko, 2008a; Ogryzko, 2009). A natural language to take these limits into account could be the formalism of quantum theory.

Intriguingly, in quantum theory, the notions of 'ensemble' and 'individual' are interrelated in a subtle way unexpected from the classical view of the world. That is, an individual object can behave in some sense as a population of objects (so called 'quantum parallelism'). This suggested to the present author that if the quantum principles are taken into account in the explanation of biological adaptation, the logic that inexorably links natural selection to populational thinking might break down. This could return an individual living organism to the 'driver's seat' of biological evolution, as the adaptive evolution could be understood as a result of 'selection in the population of virtual states' of an *individual organism* (Ogryzko, 1994; Ogryzko, 1997; Ogryzko, 2007; Ogryzko, 2008b). The most important difference of this concept from regular Darwinian mechanism is the inability to separate variations from selection (borrowing terminology from probability theory, in this case the sampling space depends on the conditions of observation, i.e. the state of environment) (Ogryzko, 1997; Ogryzko, 2008b). Other approaches have also implicated quantum theory in the phenomenon of adaptive mutations (Goswami & Todd, 1997; McFadden & Al-Khalili, 1999).

Despite continuous efforts to explain the phenomenon of adaptive mutations by special molecular mechanisms (such as a transient hypermutable state (Foster, 1998; Hall, 1991), and transient gene amplification (Pettersson ME, 2005; Roth et al., 2006)), their validity has been questioned (Seger et al., 2003; Stumpf et al., 2007). Thus, the phenomenon is still very poorly understood, keeping the chances that the deeper understanding of the physics of Life, and quantum principles in particular, would be relevant in the explanation of this biological phenomenon. In any case, I feel that the study of the Quantum Models of Adaptive Mutations (QMAMs) has the potential to develop into a field in its own right and become relevant due to the coming of age of quantum information theory and nanotechnology (Nalwa, 2004; Nielsen & Chuang, 2000). Progress in these fields could ultimately result in the realization of quantum self-reproducing automata. The question whether such devices will be able to 'cheat' the Central Dogma of molecular biology and evolve in Lamarckian, rather than Darwinian, fashion presents an independent interest, regardless of whether the regular 'earth' organisms are 'quantum self-reproducing automata' or not. Also, I hope that the concept of quantum adaptation (Ogryzko, 1997), wherein, unlike the Darwinian adaptation scheme, the 'variation' and 'selection' steps cannot be separated, could provide a bridge between the physicalist world-view and the notion of values (Ogryzko, 1994), pertinent in understanding the phenomenon of intentionality (Brentano, 1973; Chalmers, 2002; Stapp, 1999), one of the crucial subjects of the philosophy of mind.

The approach of Goswami starts with introducing the concept of consciousness in the description of the cell (Goswami & Todd, 1997), and thus merits a separate discussion outside of the scope of this paper. Here I will compare my approach (Ogryzko, 1994; Ogryzko, 1997; Ogryzko, 2007; Ogryzko, 2008b) with the approach of McFadden (and Al-Khalili) (McFadden & Al-Khalili, 1999), which appeared later in the same journal and then in a book (McFadden, 2000). I introduce language of **R-error** and **D-error** for the description of adaptive mutations and show that the first model of McFadden (McFadden & Al-Khalili, 1999) cannot be considered as a QMAM. I further introduce the hypothesis of '**R-D-error** correlation' and compare how my and McFadden's approaches fare in the context of this hypothesis. I further argue that environmentally induced decoherence (**EID**) can play a positive role in preparing the state of the cell in superposition, and that the interpretation of unitary intracellular dynamics, induced by the einselection, as 'exponential reproduction in imaginary time' can help to understand the origin of the **R-D-error** correlation.

## 2. Q-cell and Q-genome approaches

For detailed description of the adaptive mutation phenomenon and of the two approaches to be compared, the reader is referred to the reviews and the original publications (Foster, 1998; Hall, 1998; McFadden & Al-Khalili, 1999; Ogryzko, 1997; Roth et al., 2006). Here, I emphasize the most relevant points.

Admittedly simplifying the real situation, the crucial observation in the phenomenon of adaptive mutations can be summarized in three sentences: 1. These mutations do not occur when the cells are kept in conditions that do not permit growth of the mutant cells (as shown in experiments with delayed application of lactose in the original experiments by Cairns (Cairns et al., 1988)). 2. Only when conditions that are permissive for the growth of the mutants are created (e.g. application of lactose) do the mutant colonies start to accumulate on the plate. 3. The mutations occur only in the genes under selection.

How can quantum theory be used here? Obviously, because quantum mechanics is a fundamental theory for description of physical reality, any explanation of adaptive mutations (even based on 'straight' molecular biology) could eventually be described at the fundamental quantum mechanical level, whatever the cost and complexity of this exercise might entail. What sets the QMAMs apart is their reliance on characteristic features of the quantum mechanical description of the world, such as entanglement or coherence (Horodecki et al., 1996; Ono & Fujikawa, 1998).

Both my and McFadden's approaches use the same general scheme to account for the main observation (Figure 1), which will be called 'fluctuation trapping' (McFadden & Al-Khalili, 2001; Ogryzko, 2007; Ogryzko, 2008b). (a) The system under consideration fluctuates reversibly between different states (W and M). In the absence of the selective substrate (e.g. lactose), the M and W states are indistinguishable by the environment, and this situation is stable, i.e., not changing with time. (b) The fluctuating state is destabilized by the application of the selective substrate (lactose), as in these conditions the M state can lead to generation of the mutant colony. As time proceeds, more individual cells on the plate get a chance to be in the M state and to be trapped due to the irreversible amplification, leading to the continuous accumulation of mutant colonies on the plate.

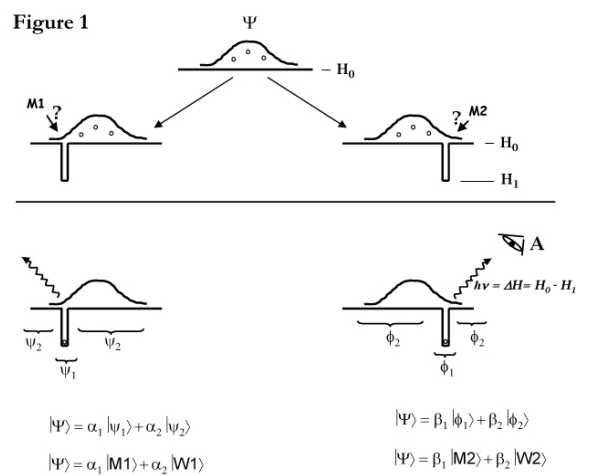

Figure 1

Both our approaches endeavor to use the formalism of quantum theory to describe the above scheme: 1. The state of reversible fluctuations corresponds to a system being in a superposition of W and M states. 2. Addition of substrate causes collapse of this superposition and corresponds to a measurement. (We can here draw an analogy with the von Neuman's Type II and I processes, respectively (Von Neumann, 1955)).

What is the nature of the fluctuation between the 'M and W' states (part (a) of the fluctuation trapping scheme)? Both models consider base tautomery – the transition of a proton from one position in a nucleotide base to another (importantly, other variations at the genetic level are also possible, and are likely to be involved in the most studied (Lac) system (Foster, 2000; Roth et al., 2006)). The base tautomery allows the same genetic sequence to be recognized in an alternative to the 'wild type' way (leading to the mutant M state which later can be trapped in the 'potential well' at the stage (b) of the proposed scheme). Importantly, although it is an important aspect of the 'W – M' fluctuation, the mere base tautomery cannot completely account for the difference between W and M states in QMAMs (see the section 3).

The two approaches differ in what is the system measured and what is the measuring device. In my approach (Ogryzko, 1997; Ogryzko, 2007; Ogryzko, 2008b), the object in the state of superposition is the bacterial cell (for a more accurate account using a density matrix, see section **6** of this paper). Correspondingly, plating of bacteria on a Petri dish (solid agar with nutrients) and waiting for the colonies to appear constitutes a measurement procedure. As bacteria stay on the Petri dish for several days, and the number of mutant colonies steadily increases with time, this measurement is somewhat similar to observation of radioactive decay (see a more detailed discussion see the end of the section **6D**).

In McFadden's approach, it is the bacterial genome (DNA) which is in the state of superposition (between alternative tautomeric forms of a particular base), and the cell itself performs measurement on DNA and thus collapses the state of DNA into a mutant one after lactose application (McFadden, 2000; McFadden & Al-Khalili, 1999; McFadden & Al-Khalili, 2001).

According to these differences, we will call the first approach (Ogryzko, 1997; Ogryzko, 2007; Ogryzko, 2008b) **Q-cell** theory and the approach of McFadden (McFadden, 2000; McFadden & Al-Khalili, 1999; McFadden & Al-Khalili, 2001) – **Q-genome** theory. Their comparison is the main subject of this paper.

### 3. The '*R-error* only' and '*D-error* only' scenarios

Here I introduce the language of ***R-error*** and ***D-error*** for the description of the fluctuation trapping model of adaptive mutations and use it to compare different scenarios to account for this phenomenon. I argue that any 'fluctuation trapping' scenario of adaptive mutations that employs only ***R-error*** or only ***D-error*** – even if it might work as a classical mechanism – cannot be considered a QMAM. Later, I will show how a particular correlation between ***R-error*** and ***D-error*** could help the bacteria to cheat the Central Dogma of Molecular Biology. Then I compare how the **Q-cell** and **Q-genome** theories can handle the implications of the ***R-D-error*** correlation.

The central question of the fluctuation trapping model is the nature of the 'mutant' M state (Fig. 1). Such a state should possess two properties in order for this model to work – its difference from the wild type W state should be both *useful* and *heritable*. '*Useful*' in this context means simply that the change in the state of the cell from W to M should enable it to consume the added substrate, whereas '*heritable*' means that this change should persevere after the cell consume the substrate and starts to proliferate. For both of these things to happen, and thus for the M state to be trapped after the substrate addition, two kinds of mutant molecules have to appear in the cell: a mutant mRNA copy of the gene (without which no active protein will be synthesized, and thus a particular genetic variation cannot be tested for its *usefulness* on the phenotypic level) and the mutant copy of the daughter strand DNA (responsible for *heritable* aspect of the phenomenon – transmission of a variation to future generations).

Either one of these mutant molecules could appear as a result of tautomery – the transition of a proton from one location of the nucleotide base to another, leading to its erroneous recognition by the transcription or replication machinery due to its mis-pairing with a wrong complementary base (Lowdin, 1965). Accordingly, by an ***R-error*** we will call a synthesis of a mutant mRNA copy of the gene due to recognition of the tautomeric form of a base by RNA-polymerase. By a ***D-error*** we will designate an analogous mistake made by DNA-polymerase.

Importantly, in this paper we consider only base tautomery as a source of the variability for the adaptive mutations (***R-*** and ***D-error***s). This is for illustrative purpose only, as other sources of genetic variability might be also involved (Foster, 2000; Roth et al., 2006). Imagine that, with a small probability, a DNA sequence can be

*reversibly* rearranged by a transposase enzyme (Berger & Haas, 2001; Rice & Baker, 2001). In this case the ***R-error*** will correspond to transcription of the rearranged DNA sequence and the ***D-error*** – to its replication.

So far, all our considerations had been relevant for any model of adaptive mutations – 'classical' or 'non-classical'. However, here is how quantum theory enters. The base tautomery can be described as a quantum mechanical superposition of proton position at two different parts of the base (Figure 2,3 top). Recognition of the base by RNA or DNA polymerase could (the issue of decoherence put aside for a moment) lead to the spreading of this superposition onto the state of the whole cell and thus play a role in the search for the adaptive mutations (McFadden & Al-Khalili, 1999; Ogryzko, 1997). How such a search could be possible using quantum principles? Here we will demonstrate that neither one of these errors alone will suffice for a 'fluctuation trapping' model to be qualified as a QMAM.

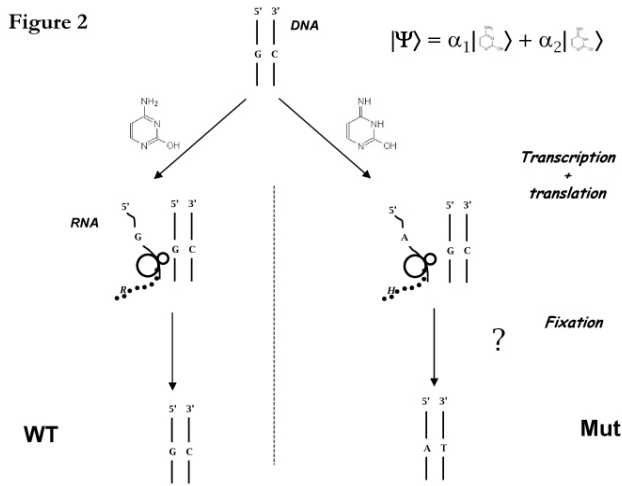

Figure 2

Consider first the '***R-error*** only' scenario (Figure 2). The generation of mutant mRNA in one of the branches will lead to appearance of an active enzyme (Figure 2, right middle), and the cell in this branch will be able to metabolize lactose. Suppose that sufficient amount of energy and building material is generated as a result of this activity, so that cell can start DNA synthesis. Since this scenario does not involve a ***D-error***, its main problem is how to generate the mutant DNA copy in order to fix the adaptive ***R-error*** for the future generations (Figure 2, right bottom). This could be done, for example, via reverse transcription (Varmus, 1987), utilizing the mutant mRNA as a template for DNA synthesis. This mechanism can in principle work, as the information about the useful ***R-error*** will be eventually transmitted to the next generation, i.e., the fluctuation will be trapped. We achieve it, however, via an introduction of an *ad hoc* mechanism (reverse transcription, which has not been demonstrated in the K-12 strain of *E.coli* used in most of the experiments on adaptive mutations (Foster, 1993)). Most importantly, no coherence, entanglement or any other quantum magic are required. Thus this model cannot qualify as a QMAM.

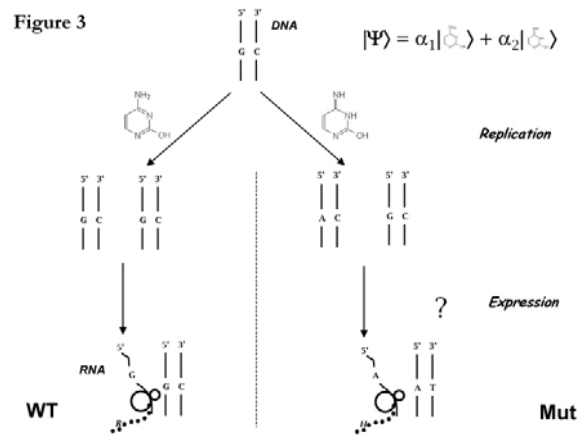

Figure 3

Now consider the '***D-error*** only' scenario (Figure 3). This scenario has a different problem to deal with: how the cell can test whether a particular ***D-error*** has a beneficial phenotypic effect (which can be done only if the mutant protein appears in the cell (Figure 3, right bottom)). Since no ***R-error*** is allowed here, the only source of the mutant mRNA and protein would be the mutant DNA copy (Figure 3, right middle). However, this implies the existence of a molecular record about the ***D-error*** in a form of a complete or partial mutant DNA copy of the gene, which could be faithfully transcribed by RNA-polymerase generating mutant mRNA (faithfulness is essential, because no ***R-error*** is allowed in this scenario). Suppose now that in the presence of lactose the cell in the M state accumulates sufficient amounts of energy and building material to resume replication (which will eventually lead to the trapping of the fluctuation). The record about the ***D-error*** will already be present in the cell in a classical form (as a mutant daughter DNA molecule), and a mutant colony will be generated. Just as in the previous case of the '***R-error*** only', no long coherence times are required for the fluctuation trapping to be accomplished. The only necessary quantum event here is the proton transition leading to the base tautomery. Thus this model also cannot qualify as a QMAM.

## 4. The McFadden and Al-Khalili (1999) model is the '*D-error* only model'

In their 1999 Biosystems paper (McFadden & Al-Khalili, 1999), McFadden and Al-Khalili claim to employ quantum coherence to explain adaptive mutations. This section shows that their model (MFAK99) corresponds to the '*D-error* only' scenario, and thus cannot be considered a QMAM.

According to this model, the proton of the nucleotide base under consideration is in a superposition of regular and tautomeric positions. The recognition of the base in this state by DNA-polymerase in the starving cell and subsequent DNA synthesis lead to the daughter DNA strand being in a superposition of mutant and wild type branches (Figure 3 and Fig.1 from reference (McFadden & Al-Khalili, 1999)). Its consequent transcription by RNA polymerase eventually leads to the cell being in superposition of two states corresponding to the cell with inactive (W) or active (M) enzyme. In the absence of substrate these states are practically undistinguishable, and are preserved in a coherent state. Addition of lactose induces fast decoherence of the M (active enzyme) branch that leads to generation of a colony. The W protons have a certain amplitude to transit to the M state and consequently be trapped by decoherence. Thus, mutant colonies are continuously generated in the course of time.

However, is coherence necessary here? Importantly, in the MFAK99 scenario, RNA-polymerase uses the mutant daughter DNA copy as a template. Suppose now that the decoherence event happens at the very first step of this scenario (before transcription), immediately after the nucleotide under the question was recognized (or mis-recognized) by DNA-polymerase, with no further superposition spreading to the state of the cell. Consider now the mutant branch in the presence of the substrate (Figure 3, right). The active enzyme in this branch will metabolize the substrate and the cell will resume replication. The mutant daughter DNA copy (which was the template for the mutant mRNA synthesis and will still be available for the DNA-polymerase) will be replicated, which will lead to the growth of a mutant colony. Therefore, the mutant will appear even though the coherence existed only for a fleeting time at the very beginning. Thus, the MFAK99 model fails to qualify as a QMAM. It includes an implicit assumption about the presence of a mutant daughter DNA strand, which, if made more explicit, renders the question of coherence in proton position irrelevant.

An alternative way to see why the MFAK99 model is not a QMAM would be to consider it as implementing a *quantum search algorithm*. The known quantum search algorithms, such as the Grover and Shor algorithms, (Grover, 1996; Shor, 1995) use two distinct features of quantum mechanics: parallelism (superposition) and interference. The parallelism allows a quantum computer to explore many possibilities at once, thus giving these algorithms much of their power. Importantly, however, to take advantage of the parallelism, an interference between different dynamic branches has to be used (Nielsen & Chuang, 2000). From this perspective, we can clearly see that only the 'parallelism' part of the quantum search algorithm has been employed in the MFAK99 model. It does not take any advantage of the interference between different branches. Thus it would not make any difference for this model if the branches were decohered from the very beginning. In principle, this model might work, but in a classical way, and it does not make sense to consider it a QMAM.

## 5. The *R-D-error* correlation

Here we will show how a combination of the *R*- and *D-error*s in one scenario would make a QMAM viable.

First, consider a scenario where both *R*- and *D-error*s are allowed. Start with RNA-polymerase and assume again that two superposed branches of the cell are created due to the base tautomery. Take the mutant branch in the presence of lactose. Again, assume that enough energy and building material is generated for some cryptic DNA replication to start. Since in this case *D-error* is allowed, mutant DNA copies can be generated with some probability during replication, and mutant colonies will eventually appear. This mechanism is not a QMAM yet, as it does not employ any exotic quantum effects[1].

Consider now a modified version of this scenario (Figure 4, right). Assume this time that there is a *correlation* between the *R*- and *D-errors*, such that the DNA-polymerase has a high probability to make exactly the same error as the error made by the RNA-polymerase (i.e., both mis-recognize the same nucleotide in the same erroneous way) (Ogryzko, 2007). We will call this scenario '*R-D-error* correlation'. This model of adaptive mutation can work more efficiently than the previous one, because, unlike in the above case, the ***D-error***s will replicate and fix exactly those genomic variations that were tested via the ***R-error***s to have a beneficial effect.

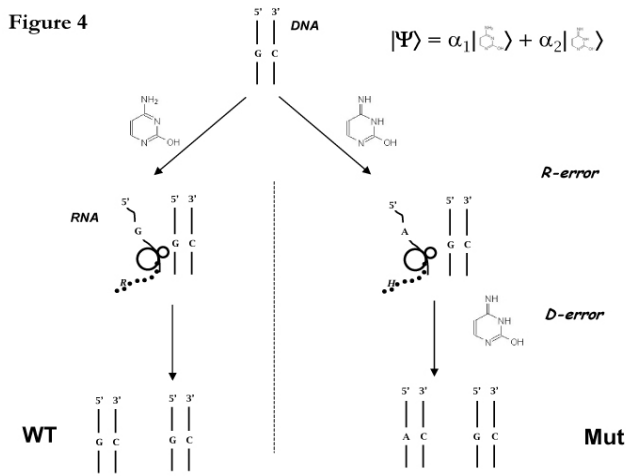

Figure 4

Certainly, no 'classical' mechanisms known to molecular biology can provide a basis for such hypothetical ***R-D-error*** correlation in the cell. On the other hand, among the main features of quantum theory are exactly the non-classical correlations between different events that cannot be accounted for by regular causal mechanisms. For example, entanglement, the most characteristic feature of quantum theory, manifests itself in correlations between the results of measurements performed on different parts of a composite system (Horodecki et al., 1996). Therefore, a 'fluctuation trapping' model of adaptive mutations, in which the M state corresponds to the correlated ***R-D-error***, could be the QMAM that we are looking for.

Two related problems immediately arise. Attempts to implement protocols that use entangled (or more generally – coherent) states in quantum computation or cryptography show that these states are sensitive to the interaction of the experimental system with its environment; they are quickly destroyed in a process aptly called environmentally induced decoherence (**EID**). Then the first problem is – how can any nonclassical correlations in the cell survive the **EID**? A related problem is – assuming that some nonclassical correlations in the cell can be somehow protected from the **EID**, why is it precisely the ***R-D-error*** correlations that will survive the decoherence?

The rest of the paper will discuss how the **Q-cell** and **Q-genome** approaches can deal with these two questions.

## 6. The Q-cell approach and the *R-D-error* correlation.

This section is divided into several parts, discussing: 1) How, contrary to a common misconception, decoherence can play a positive role in stabilizing some non-classical correlations in a macroscopic object, in particular in a living cell; 2) How the **R-D-error** correlation can be justified suggesting a new interpretation of the quantum dynamic as 'exponential growth in imaginary time'.

### 6A. Adaptation via einselection. Positive role of decoherence.

In quantum theory, the state space of a composite system **A** is a tensor product of the state spaces of its parts $A_i$:

$$A = A_1 \otimes A_2 \otimes A_3 \ldots \qquad (6A.1)$$

Accordingly, the vast majority of the possible states of every macroscopic system correspond to superpositions of the form (using Dirac's notation)

$$|\Psi\rangle = \alpha(|^1\psi_1\rangle|^1\psi_2\rangle|^1\psi_3\rangle\ldots) + \beta(|^2\psi_1\rangle|^2\psi_2\rangle|^2\psi_3\rangle\ldots) +$$
$$+ \gamma(|^3\psi_1\rangle|^3\psi_2\rangle|^3\psi_3\rangle\ldots) +\ldots \qquad (6A.2)$$

where $|^i\psi_j\rangle$ are different states (i) of its many different parts (j). Most of the $|\Psi\rangle$, similarly to the notorious Schroedinger cat, are never observed and have to be explained away. However, some such states of the composite system are absolutely legitimate and reflect interactions between different parts of the system holding it together (for example, covalent bonds). An explanation of the transition from the quantum to the classical world, developed by Zeh, Zurek and their collaborators (Zeh, 1970; Zurek, 2003; Zurek et al., 1993), separates one class of $|\Psi\rangle$ states from another by employing the environmentally induced decoherence (**EID**) both in a destructive and in a constructive way. Namely, **EID** will suppress most of the exotic cat-like states, but some of the states, called in this context 'preferred states', will be selected and stabilized by **EID**.

In this approach, the physical system is described by a reduced density matrix $\rho^s$, obtained from the density matrix $\rho$ of the total system **S+E**

(including system **S** coupled to its environment **E**),

$$\rho = |\Psi_{ES}\rangle\langle\Psi_{ES}| \tag{6A.3}$$

by tracing out the environmental degrees of freedom:

$$\rho^s = Tr_E|\Psi_{ES}\rangle\langle\Psi_{ES}| \tag{6A.4}$$

Starting from an arbitrary state of the joint system **(S+E)**, and choosing some basis for a description, the reduced density matrix of **S**:

$$\rho^s = \Sigma \alpha_i \alpha^*_j \langle \varepsilon_i | \varepsilon_j \rangle |s_i\rangle\langle s_j| \tag{6A.5}$$

will in general contain off-diagonal terms $|s_i\rangle\langle s_j|$. These terms (also called *coherences*) correspond to interference between the basis states and are responsible for quantum effects. *Decoherence* refers to the fact that these off-diagonal terms will often quickly vanish with time (their contributions will average out to zero), as the dynamic evolution of the joint system (S+E) will generally lead to rapid separation between the different basis states of S, due to their entanglement with the uncontrollable environment. The $\rho^s$ becomes diagonal, and the ensuing absence of interference between different basis states is proposed to explain why macroscopic superposition states (such as Schroedinger cat) can never be observed.

Importantly, decoherence worked here because we chose the right basis (the one that would allow evolution to a diagonal form). The vanishing of the off-diagonal terms in a particular basis is usually justified by the nature of the coupling between a system and its environment (i.e., whether it can distinguish between different states of the system). In particular, if the **E-S** interaction is position dependent and the environment can be approximated by a thermal bath, it is the position basis that will allow diagonalization and thus survive decoherence (Zurek, 2003). On the other hand, if the interaction Hamiltonian is periodic in position, decoherence will lead to the momentum basis as the preferred one. In the general case, when the exact form of the preferred basis cannot be easily determined, the **EID** approach provides a formal criterion for state survival, based on its commutativity with the **E-S** interaction Hamiltonian (Zurek, 2003). This criterion is at the heart of environmentally induced superselection (einselection), which will be used in our general description of the adaptation process.

The possibility of using einselection for the description of biological adaptation was proposed in my previous publication (Ogryzko, 1997), and here it will be considered in more detail. In order to accommodate the einselection scheme to biology, we first must acknowledge an important difference in how the concept of environment is used in biology compared to physics. Whereas in physics it commonly plays a role of a homogenous background, the environment of biologists is far more interesting. First, usually it is at least as ordered as the organism itself (for example, it can contain molecules of various structures that can be utilized by a cell). Second, the environment varies. Furthermore, the specific and subtle relations between the living things and their changing environment is, in fact, one of the main subjects of the life sciences, the 'bread and butter' of biologists.

Accordingly, the application of the idea of **EID** to a biological system will generally require to consider several different environments: $E_0$, $E_1$, $E_2$, $E_3$…. We can formally write that each $E_i$ will select its own set ($^i s_i$) of preferred states of the same system via the **EID** mechanism:

$$\begin{aligned}&E_0 : (^1s_0, {}^2s_0, {}^3s_0…),\\&E_1 : (^1s_1, {}^2s_1, {}^3s_1…),\\&E_2 : (^1s_2, {}^2s_2, {}^3s_2…),\\&E_3 : (^1s_3, {}^2s_3, {}^3s_3…),\end{aligned} \tag{6A.6}$$

The exact form of these states generally cannot be known, as the ordered character of the environment (such as presence or absence of various substrates) makes the procedure of tracing out the environmental degrees of freedom far from trivial. The only thing that should concern us, however, is that formally the preferred states will be determined by the symmetries of the interaction Hamiltonian, and in general their spectrum will be different for each environment.

Consider now the system in the environment $E_0$. It will be described by a reduced density matrix $\rho_0$, reflecting uncertainty in its state due to the interaction with environment.

What happens if we change an environment to $E_1$? The new environment will select a different spectrum of preferred states. The original state $\rho^s_0$ cannot be in general represented by a diagonal density matrix in the new preferred basis corresponding to the $E_1$. Some off-diagonal terms $|^i s_1\rangle\langle^j s_1|$ will have to be present in the new description of $\rho^s_0$. Those $|^i s_1\rangle\langle^j s_1|$ will duly vanish in the new conditions via **EID**, describing an adaptation of the system to the new environment and emergence of new state $\rho^s_1$.

$$\langle {}^i\varepsilon_1 | {}^j\varepsilon_1 \rangle \to 0, \quad \rho^s_0 \to \rho^s_1 \quad (6A.7)$$

Importantly, however, if we consider the system in its previous state $\rho^s_0$ before the actual change in environment from $E_0$ to $E_1$, we will have to admit that the coherence represented by these off-diagonal terms $|{}^is_1\rangle\langle {}^js_1|$ *was present all along* in the system when it still was in environment $E_0$. In fact, this coherence was stabilized and the $|{}^is_1\rangle\langle {}^js_1|$ presence was ensured by **EID**.

We thus obtain very general and economical description of the adaptation process, where **EID** plays a dual role - it stabilizes (prepares) certain coherent states in a particular environment and destabilizes the very same states in other environments. I find this description very satisfying, due to the following features: 1) It views both biological and physical phenomena of adaptation from a unified 'selectionist' perspective, 2) It gives due weight to the more important and subtle role of environment in the case of biological systems, 3) It shows how, instead of being an obstacle, decoherence can be a positive force on both stages of adaptation – before and after change of environment.

### 6B. Properties of the starving state of the cell as an einselected state.

Applying the above general description to the phenomenon of adaptive mutations, I propose to consider a bacterial cell in the absence of substrate (starving cell) to be in a state einselected in this environment (($E_0$) as discussed in 6A). This state will be referred to henceforth as a **U** state[2]. In this section I will consider two consequences of this proposal.

First, this suggestion provides more legitimacy to the statement 'cell in a state of superposition of mutant and wild type states', central to the **Q-cell** approach. We have to make clear distinction between two types of superposed states of a macroscopic object: 1). a superposition of distinct macroscopic states and 2). a more general idea of a macroscopic object being in a state of superposition of some eigenstates of a particular operator. An example of the former is the Schroedinger cat, which is very counter-intuitive and hard to come by. The example of the latter is phonon in a crystal lattice – phonon is usually delocalized in the lattice, therefore, its state can be represented as a superposition of the eigenstates of the position operator. However, talking about phonon we are in fact describing the dynamics of the lattice itself (phonon is a quasiparticle). Thus it is the crystal lattice (macroscopic object) that is described using the concept of superposition; and compared to the exotic Schroedinger cat, phonon is an everyday occurrence.

As the discussion in the section 6A indicates, the proposed 'superposition state of cell' is of the second kind, since the wild type and mutant states of the cell are proposed to be indistinguishable from each other in the conditions of starvation ($E_0$), both being components of the einselected state **U**. This notion of superposition challenges neither common sense nor observation and merely describes the potential existence of several outcomes of a cell's interaction with a different environment (such as $E_1$). Only after the cell is put in this new environment ($E_0 \to E_1$), which can distinguish between the wild type and mutant cells, does the 'superposition state' becomes unstable and destroyed by **EID**[3]. (See Figure 5 for an illustration of this idea on the example of a crystal lattice). This description, suggested previously (Ogryzko, 1997; Ogryzko, 2007), implies an existence of an operator $O_{E1}$ acting on the Hilbert space of the states of the cell, such that the M and W states are the eigenstates of this operator[4]. Another operator (call it $O_{E0}$) corresponds to the old environment $E_0$ with the **U** state being its eigenstate (and the fact that **U** is represented as a superposition of M and W implies that these two operators do not commute).

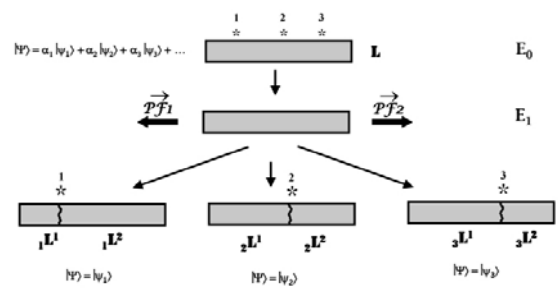

Figure 5

A second consequence of the above proposal concerns the physical nature of intracellular processes in the starving cells. Adaptive mutations do occur in starving cells, therefore the gene expression and DNA replication machineries should be exhibiting some level of activity, i.e., some molecular processes have to take place in it.

How is it possible to reconcile this point with the seemingly static nature of the idea of 'einselected' or 'preferred' state? First, one can argue that in quantum mechanics a stationary state can be also considered as dynamic, insofar as with time the state vector describing it changes its phase:

$$|\psi(0)\rangle \rightarrow |\psi(t)\rangle = e^{-\phi tH}|\psi(0)\rangle \qquad (6B.1)$$

Furthermore, if a different basis for the description of the system is chosen, the dynamics will appear more sophisticated, and will also include transitions between different components of the basis. This alternative description (**MB basis**) will be discussed in more detail later (6.C.2), as relevant for the description of molecular processes taking place in the starving cells. Regardless of basis choice, however, the very definition of an einselected state requires that its dynamics is protected from decoherence, therefore the molecular processes in the starving cell have to be described by unitary dynamics[5] (von Neuman II process (Von Neumann, 1955)).

Importantly, the notion of a unitary nature of intracellular dynamics in the starving cell is not an additional independent suggestion. It is a consequence of the need to reconcile the proposal of the starving cell being in an einselected state (**U** state) with the fact that some molecular processes do happen in it. Nevertheless, it has far-reaching implications. In particular, it challenges the conventional wisdom of the irreversibility of intracellular processes. The discussion of all ramifications of this idea is beyond the scope of the present paper (Ogryzko, 2009). We can briefly state, however, that the irreversibility of intracellular dynamics on a larger time scale is not ruled out by this proposal. We merely suggest separating intracellular processes into two different classes. The first class corresponds to the preferred states, protected from decoherence due to einselection and undergoing unitary-type evolution (von Neuman II class process). At a larger time scale, however, these states will reveal their *metastable* character, and the description of intracellular dynamics will have to be supplemented by the second-class processes – those mostly represent transitions between the preferred states of different classes (e.g., **E₀** to **E₁** to **E₂**), caused by the changes in environment (von Neuman I class). The proposal to consider the starving cell to be in a **U** state (einselected in the substrate-free environment) implies that for the time scales relevant for the adaptive mutations, the intracellular dynamics can be considered as unitary.

The hypothesis of a unitary character of intracellular dynamics in the starving cell provides us with the first step in the justification of the proposed ***R-D-error*** correlation. In a unitary process no information can be lost (Nielsen & Chuang, 2000). Therefore, if an ***R-error*** takes place in the starving cell, the cell will be able to keep track of it, i.e., the information about the cause of the appearance of a mutant protein will be preserved in the state of the cell. Furthermore, this memory can have an effect on the probability of a ***D-error*** happening in the same cell. To better grasp this idea, we need to change our perspective and depart from biochemical intuition that relies on experiments *in vitro*, shifting instead to consideration of enzymatic events as they happen in the context of an individual living cell. The difference between *in vitro* and *in vivo* cases is essential. When the DNA polymerase reaction is modeled *in vitro* by adding the enzyme and substrates to each other, these components can be safely considered to be separable, as they were prepared independently from each other before the interaction. The outcome of this interaction (the sequence of the daughter DNA molecule) will be fairly consistent with the *in vitro* measured value of the difference in free energy ($\Delta G = G_r - G_t$) between the regular and tautomeric forms of the nucleotide base, responsible for a certain probability of a ***D-error*** *in vitro*. However, the *in vivo* situation is different. The correct physical description should include all interacting components (DNA-polymerase, nucleotide precursor and the DNA template) as parts of the bigger system (starving cell) undergoing unitary evolution. In this description, the parts *a priori* cannot be taken as independent from each other, neither before nor after interaction. Therefore, when considered *in vivo*, the outcome of the interaction will be determined by the state of the whole cell, and thus ultimately ***D-error*** can depend on an ***R-error*** happening in the same cell.

Thus, to briefly summarize the first step in the justification of the ***R-D-error*** correlation, the possibility of a nonclassical correlation in a starving cell arises due to einselection that imposes the requirement of unitarity on the intracellular dynamics. The unitary nature of the dynamics allows the cell to keep track of the useful ***R-error***. In turn, this memory can affect the DNA-polymerase interaction with its substrates *in vivo*, leading eventually to a correlation between the ***R-*** and ***D-*** errors.

All that said, we are certainly not out of the woods yet. The fact that einselection could lead to a correlation between the actions of RNA- and DNA-polymerases does not by itself guarantee that the results of the *D-error* will be skewed exactly in the way favoring the adaptive mutations. Imagine the following constraint imposed by einselection – whenever RNA-polymerase makes an error, DNA-polymerase always recognizes the same nucleotide base in a correct way, and vice versa:

$$P = \{(R_{er}, D_{cor}), (R_{cor}, D_{er})\} \quad \quad (6B.2)$$

where P is the set of possible outcomes, consisting of two elements: $(R_{er}, D_{cor})$, corresponding to combination of *R-error* and no *D-error* and $(R_{er}, D_{cor})$, corresponding to combination of *D-error* and no *R-error*.

In this hypothetical scenario some sort of correlation between the two events is clearly present. However, it is not the *R-D-error* correlation that we need, as it does not help to fix the adaptive mutation. The required correlation would have the form:

$$P' = \{(R_{er}, D_{er}), (R_{cor}, D_{cor})\}, \quad \quad (6B.3)$$

In other words, we still do not have the answer to the second question – why is it precisely the '*R-D-error* correlated' states (yielding **6B.3)** that will be selected by **EID** out of the vast number of potential states of the cell inhabiting the Hilbert space of our system?

The next section will suggest a new interpretation of unitary dynamics as 'exponential growth in imaginary time'. According to this interpretation, the *R-D-error* correlation will follow from the very fact that cell is able to self-reproduce.

*6C. Exponential growth in imaginary time.*

**6C.1**

It has long been noted that the Schroedinger equation, which describes unitary dynamics[6]:

$$i\hbar(\partial/\partial t)\Psi = -(\hbar^2/2m)\nabla^2\Psi \quad \quad (6C.1)$$

can be understood as a heat (diffusion) equation:

$$(\partial/\partial t)\Psi = D\nabla^2\Psi, \quad D > 0 \quad \quad (6C.2)$$

occurring in imaginary time *it*, instead of real time *t* (Fenyes, 1952; Nelson, 1966). The physical meaning of this interpretation is unclear. Nevertheless, it illustrates an important point – the requirement to have unitary character imposes strict constraints on the dynamics of a system under consideration. The change from real time *t* to imaginary time *it* (the so called Wick rotation) turns an irreversible and unidirectional process of redistribution of a physical system in its state space (diffusion, mathematically described by a semigroup) into a reversible deterministic process describing the (oscillating) dynamics of a standing wave in the high-dimensional state space of the system (essentially, a state where, if the basis is chosen correctly, 'nothing happens' except for a phase change; the process is mathematically described by a group). In the case of a composite system, one can see these constraints as reflecting the nonlocal character of unitary dynamics[7]. For, although the parts of a composite system (*a* + *b*) exhibit loss of coherence with time:

$$\rho^a(t_0) = Tr_b|\Psi_{AB}\rangle\langle\Psi_{AB}| = \Sigma\alpha_i\alpha^*_j\langle b_i|b_j\rangle|a_i\rangle\langle a_j| \rightarrow$$
$$\rightarrow \rho^a(t) = \Sigma\alpha^2_i |a_i\rangle\langle a_i|$$

$$\rho^b(t_0) = Tr_a|\Psi_{AB}\rangle\langle\Psi_{AB}| = \Sigma\beta_i\beta^*_j\langle a_i|a_j\rangle|b_i\rangle\langle b_j| \rightarrow$$
$$\rightarrow \rho^b(t) = \Sigma\beta^2_i |b_i\rangle\langle b_i| \quad \quad (6C.3)$$

(notice the loss of the off–diagonals in both cases), the unitary character of the dynamics of the joint system (*a+b*) dictates that the parts *a* and *b* evolve in a correlated way such that there is no irreversible deterioration of the state of the whole system (*a+b*), i.e., the overall dynamics is conservative (the information about the states of the parts has been converted into information about correlation in their behavior (Horodecki & Horodecki, 1998)).

The '*diffusion*' interpretation of the Schroedinger equation gives no clear physical meaning to Wick rotation, simply using it as a formal mathematical trick. This is somewhat of a drawback and motivates us to take the following crucial step. As long as we are considering the dynamics of the cell in the **U** state as an analytic continuation of some 'real time' process (that is, diffusion) to the imaginary coordinate, we might equally consider it as the 'imaginary time' counterpart of a different 'real time' process, namely, *copying*. We will also interchangeably use other terms, such as 'cloning', 'exponential growth' and 'reproduction'. As will be argued below, despite the dramatic differences in their meaning and behavior as real time processes (described by positive exponential $e^{-t}$ and negative one $e^{+t}$, correspondingly), the formal description of both exponential growth and diffusion looks exactly the same – like stationary waves (harmonic oscillations) – when the real time coordinate *t* is replaced to an imaginary coordinate *it*. However, the 'exponential growth/copying' interpretation has the following advantages in our case:

a. It is suited especially well for the description of intracellular dynamics in the **U** state, as the cells possess all prerequisites (including enzymatic machinery and genetic information) required for self-reproduction – as attested by the very empirical fact that they can proliferate.

b. The new interpretation of unitary dynamics naturally provides us with a procedure to *implement Wick rotation in the real world* – we can convert an 'imaginary time' reproduction into a 'real time' one by simply adding a substrate and thus letting the cell proliferate.

Before describing how this proposal could help with the justification of the ***R-D-error*** correlation (**6D**), I will consider local and global aspects of the proposed interpretation, and also its relation to the no-cloning theorem (Dieks, 1982; Wootters & Zurek, 1982).

**6C.2. Global and local aspects of the suggested interpretation. MB basis versus PR basis.**

An important feature of the suggested proposal is the assumption that most of enzymatic events that occur during regular cell growth also take place in the **U** state (i.e., the starving cell is 'reproducing itself in imaginary time'). However, consistent with the above discussion (eq **(6C.3)**), the unitary character of the intracellular dynamics in the einselected state entails existence of correlations between the actions of different enzymes in the cell, such that the overall dynamics of the cell in the **U** state is physically conservative (dissipation- and decoherence-free), and hence preserves all information about the state of the system.

The conservative nature of unitary dynamics is most obviously seen if we choose a basis for its description that corresponds to a 'simple' phase rotation (we shall call it the **PRB** basis):

$$|\psi(t)\rangle = e^{-\phi tH}|\psi(0)\rangle \qquad (6C.4)$$

In this description, 'nothing happens' except phase rotation, so the **U** state can be considered as static. In accordance with the QM formalism, other descriptions of the same **U** state are also possible. More in agreement with the molecular biological intuition is the basis that we will call **MBB** (for **M**olecular **B**iology **B**asis). The elements of the **MBB** specify locations of every nucleus and electron in the cell, i.e., they carry the structural information about molecules, their position and orientation in the cell. In this basis the intracellular dynamics is described by Laplacian operator ($\nabla^2$) that relates the rate of change in the occupation of a particular state A ($d\psi/dt$) with the local situation in its neighborhood. Usually this dynamics is interpreted as describing transitions between different **MBB** states, due to two main factors: a) enzymatic activity, accounting for covalent bond rearrangements, active transport, etc.; and b) diffusion, responsible for passive changes in location and orientation of molecules in the cell. In this interpretation, the overall dynamics can be understood as generalized diffusion (random walk) in high-dimensional space of the states of cell (Welch, 1992). Clearly, the new 'copying' interpretation of unitary dynamics proposed here will also require an alternative justification of the use of the Laplacian operator. However this task is beyond the scope of the present article (also see the discussion of the Euclidean approach to the reproduction problem in 8.4).

Two aspects of the relationship between the **PRB** and the **MBB** should be emphasized here:

***a. Connectivity***. Two elements of the **MBB** *a* and *b* will be called *connected* (*a* ~ *b*) if state *a* can be reached from state *b* by a path that includes intermediate states *c*, *d*, … and transitions (enzymatic acts and diffusion) between the states involved in the path. This property is transitive (if *a* ~ *b* and *b* ~ *c*, then *a* ~ *c*) and, due to the reversibility of unitary dynamics, reflective (if *a* ~ *b*, then *b* ~ *a*). Since the **PRB** states are the stationary solutions of the dynamic equations, they should be naturally closed in respect to connectivity, i.e., if a **PRB** state X includes an **MBB** state *a*, all **MBB** states $b_i$ ~ *a* must also be included in state X.

***b. Complex coefficients.*** In general, the **MBB** states enter into the expansion of the **PRB** state with complex coefficients, reflecting the fact that in quantum theory the state of an object is described by amplitudes and not probabilities[8]. Accordingly, the density matrix describing the **U** state using the **MBB** will also contain off-diagonal elements that are complex numbers. These off-diagonals reflect interference between different elements of the basis, and as the discussion in the 6.A indicates, they are responsible for stability of the **U** state.

### 6C.3. Role of the non-cloning theorem.

How the idea of unitary evolution as 'reproduction in imaginary time' is consistent with the non-cloning theorem, which forbids copying of arbitrary quantum states (Dieks, 1982; Wootters & Zurek, 1982)? One can consider two ways to make these two notions compatible. First, one can notice that cloning in 'real time' is not always forbidden, but in fact, it is allowed with respect to some orthogonal basis (see Appendix). Then, the basis elements of the diagonalized density matrix describing the einselected state **U** could be the orthogonal states replicating in 'imaginary time' without violating the non-cloning theorem. Alternatively, one can argue that since the starving cell in the **U** state does not undergo actual replication and no external substrate is consumed, the arguments that forbid the copying of arbitrary quantum states do not apply to the case of 'reproduction in imaginary time'. This interesting possibility is further considered in the **Appendix**.

It is beyond the scope of the present paper to develop a unified description of the reproduction process applicable to both real and imaginary time[9]. In any case, the need to describe the transition from 'imaginary time' to 'real time' replication after substrate addition requires that, additionally to the **PR** and **MB** bases, we have to introduce a third basis for the decomposition of the **U** state, which will be called the 'cloning basis' or **CB$_E$**. As its elements, **CB$_E$** contains the states of the cell that can be cloned in real time in particular environment **E**. According to the non-cloning theorem, the elements of this basis have to be orthogonal to each other. We know from our experience that once a cell has produced a colony, its genome can be extracted and its sequence be determined with an arbitrary precision. Thus, the states of the cells with different genomes seem to naturally qualify as the elements of this basis. However, the situation is more subtle, as the next section (specifically, the comment 1 at its end) will show[10].

Very importantly and bearing with the discussion from 6.1, the **U** states are not the elements of the cloning basis **CB$_E$**. As one of the reasons for this, we can consider the effect of base tautomery on the state of the cell. The transition of a proton to an alternative position in a particular nucleotide will lead to appearance of the state of the cell containing copies of DNA and mRNA carrying mutations in this position. According to the connectivity property (6.3.1.a), these states will have to contribute into the same **U** state. However, they correspond to a different element of the **CB$_E$**. Thus, generally, a **U** state will have to be represented as a linear combination of several elements of the cloning basis (with the wild type component being predominant).

### 6D. Justification of the *R-D-error* correlation

Now we are ready to proceed further with the justification of the ***R-D-error*** correlation in the framework of the **Q-cell** theory. As already mentioned before, we consider base tautomery as the sole source of genetic variability. However, this is for illustrative purposes only, as other types of variation at the genetic level are quite possible and most likely play a role in adaptive mutagenesis (Foster, 2000; Roth et al., 2006).

Consider a starving cell first. We will focus on the tautomery of the nucleotide base that plays a role in adaptive mutation from Lac- to the Lac+. Bearing with the previous consideration of the effect of base tautomery on the **U** state, both mutant and wild type DNA and mRNA molecules can be present in the **U** state (i.e., there will be a small probability of observing these mutant molecules in an individual cell). Thus, in the **MB** basis the density matrix describing the **U** state can be written as follows (only the terms relevant for our discussion are shown):

$$\begin{bmatrix} r_m r^*_m, & r_m r^*_w, & r_m d^*_m, & r_m d^*_w, & r_m \ldots, \ldots \\ r_w r^*_m, & r_w r^*_w, & r_w d^*_m, & r_w d^*_w, & r_w \ldots, \ldots \\ d_m r^*_m, & d_m r^*_w, & d_m d^*_m, & d_m d^*_w, & d_m \ldots, \ldots \\ d_w r^*_m, & d_w r^*_w, & d_w d^*_m, & d_w d^*_w, & d_w \ldots, \ldots \\ \ldots, & \ldots, & \ldots, & \ldots, & \ldots, \ldots \end{bmatrix} \quad (6D.1)$$

where the elements of the matrix describe the contributions of different **MBB** states and their interference with each other. Namely, the term $r_m r^*_m$ corresponds to the contribution of $|R_m\rangle$, whereas $d_m r^*_w$ corresponds to the interference between the $|D_m\rangle$ and $|R_w\rangle$, etc. Here, the $|R_w\rangle$ and $|D_w\rangle$ are the states of the cell containing wild type mRNA or DNA copies of genome, and $|R_m\rangle$, $|D_m\rangle$ are states of cell containing mutant form of mRNA or DNA copies of genome.

The off-diagonal terms ($r_m r^*_w$, $d_m r^*_w$, … etc) represent interference between the different states of the **MBB** contributing to the **U** state. It is important to classify these off-diagonal terms to two types: the first type corresponds to the interference between the wild and mutant type states (such as $r_m r^*_w$, $r_m d^*_w$, $r_w d^*_m$, $d_w d^*_m$, ...; we will call them WM off-diagonals), and the second type corresponds to the interference between the states that contain mRNA and DNA copies of the same (wild or mutant) forms of DNA ($r_w d^*_w$, $r_m d^*_m$, ...; we will call them RD off-diagonals).

As discussed in 6C.1.b, the presence of both types of off-diagonal terms is important for preserving the unitary character of the intracellular dynamics in the **U** state, that is for keeping it stable. On the one hand, the WM off-diagonals keep under control the effects of recognition errors due to base tautomery. These errors would have a discernable effect in the 'real time' proliferation regime (due to their irreversible amplification), but have to be tolerated in the regime of 'imaginary time' proliferation (where 'nothing happens'). On the other hand, regardless of any tautomery, the action of DNA- and RNA- polymerases in 'real time' would erode the system's state, as it would irreversibly consume the cellular resources and dissipate energy. The RD off-diagonals, which correspond to the interference between the states of the cell containing DNA or mRNA copies of the Lac gene, are responsible for making the action of these enzymes compatible with the unitary nature of the dynamics of the **U** state.

After discussing the structure of the density matrix describing the starving cell (the **U** state), let's consider a change in its environment (**E$_0$ → E$_1$**), allowing our cell to proliferate. First we consider addition of glucose, a *generic* substrate that allows proliferation of both wild type (Cm) and mutant (Cw) variants of the cell. Keeping with the suggested interpretation (6C.b), the addition of substrate to a starving cell followed by cell reproduction is described by Wick rotation, which converts the 'imaginary time' replication regime to the 'real-time' replication regime. According to the arguments from the non-cloning theorem (Dieks, 1982; Wootters & Zurek, 1982), the superposition of the Cm and Cw elements of the **CB$_E$** (corresponding to the mutant and wild type states) cannot be amplified. This is consistent with the empirical fact that the resulting colony can only correspond to either wild type or to a mutant (with the probability to obtain the wild type colony being significantly higher than that of the mutant one).

Consider now a different change in environment (**E$_0$ → E$_2$**) – addition of lactose, a *specific* substrate that allows only the mutant cell (Cm) to proliferate. In our description, this situation will correspond to Wick rotation happening only for the Cm component of the **U** state, since only this component can generate a colony in these conditions.

Importantly, in our description, the fates of the off-diagonal terms of the density matrix are different after the change in environment and ensuing Wick rotation. The WM off-diagonals have to vanish as a part of decoherence process, because substrate addition will make the Cm and Cw elements of the **CB$_E$** basis distinguishable by environment. Importantly, this is not the case for the RD off-diagonals. The $|R_w\rangle$ and $|D_w\rangle$ states **(**or $|R_m\rangle$ and $|D_m\rangle$ states**)** belong to the same element of the **CB$_E$** basis (Cw or Cm, correspondingly), and nothing in our formalism suggests that the terms for $|R_w\rangle\langle D_w|$ and $|R_m\rangle\langle D_m|$ also have to disappear after the Wick rotation. The crucial idea here is that the change in environment will not make the states containing the wild type mRNA and DNA molecules (or mutant ones) distinguishable and thus no decoherence between them will ensue.

Finally, consider the Cm component. As a part of the **U** state, it contained the mutant versions of the DNA and mRNA molecules. The fact that it underwent Wick rotation and was amplified after lactose addition indicates that in this case both DNA- and RNA-polymerases mis-recognized the same base, i.e., *there was a correlation between the **R**- and **D**-errors*.

To summarize, the interpretation of the dynamics of **U** state as 'reproduction in imaginary time' allows us to justify the ***R-D-error*** correlation by suggesting that the **U** state can be represented as a superposition of two components: a wild type and a mutant one (Cw and Cm), both undergoing reproduction in imaginary time (see the comment 1 at the end of this section). Since both RNA-polymerase and DNA-polymerase have to be involved in the imaginary time reproduction of the mutant component of the **U** state, this entails that both enzymes mis-recognize the same base, generating mutant RNA and DNA copies of the genome of the cell. While the cell is in the **U** state, the appearance of these mutant molecules (as well as their wild type counterparts) can be nothing other than a reversible fluctuation of the state of the cell (corresponding to what has previously been called 'virtual mutation' (Ogryzko, 2007)). However, an addition of a substrate that allows the mutant component Cm to proliferate will lead to the irreversible amplification of this particular fluctuation.

Two final comments are in order.

1. The first comment concerns the structure of the $CB_E$ basis and its relationship to particular environment. While we are describing the cell in the **U** state, every nucleotide base in DNA able to tautomerise will contribute to the uncertainty of the **U** state due to generation of various mutant RNA and DNA copies. No particular nucleotide position plays any special role in this case. The addition of a *generic* substrate that is permissive for growth of all variants (such as glucose, $E_0 \rightarrow E_1$) will allow amplification of every mutant resulting from this uncertainty[11]. DNA can be extracted from the resulting colonies, and its sequence can be determined with an arbitrary precision. Therefore, the $CB_{E1}$ basis has to contain every genetic variation (resulting from the base tautomery) as a separate element, so that the state of the cell can be expanded as

$$|\Psi\rangle = c_w|\psi_w\rangle + \Sigma c_i|\psi_i\rangle,$$

where '**w**' labels wild type state, '**i**' labels all possible mutant states, and $c_w \gg c_i$. However, if we now add a *selective* substrate (such as lactose, described here as a different change of environment, $E_0 \rightarrow E_2$), this symmetry between different genomic positions breaks down. Those few variants ($|\psi_l\rangle$) that are capable to grow on lactose will lead to colony growth, whereas the wild type $|\psi_w\rangle$ and the majority of remaining variants $|\psi_i\rangle$ (unable to grow on lactose) will remain un-amplified, and thus will be undistinguishable from each other. Insofar as there remains an uncertainty as to its actual sequence, the wild type state together with all variant states unable to amplify will constitute one element of the $CB_{E2}$ basis. Thus, the cloning basis that we have to use to expand the **U** state (the components that can or cannot grow, Cm and Cw components, respectively) depends on the particular environment. As has been pointed out previously, this means that the spectrum of variations cannot be separated from selection in this adaptation scheme (borrowing terminology from probability theory, the sampling space is determined by the conditions of observation), which is the point of principal departure from the canonical Darwinian selection scheme (Ogryzko, 1994; Ogryzko, 1997; Ogryzko, 2008b).

2. The second comment concerns the kinetics of appearance of the mutant colonies on the Petri dish. As argued in the section 6B, the substrate addition can be considered as a change to a new environment: $E_0 \rightarrow E_2$ that can distinguish between the mutant and wild type states of the cell. Another way to formulate the same idea is to say that the new environment $E_2$ suppresses the interference between the Cw and Cm states, effectively generating superselection rules (SSR) that forbid the transition between these states[12]. Importantly, however, due to the fact that an individual cell is a finite physical system, these SSR are not absolute and thus remain permissive for some transition between the Cw and Cm states. The remaining possibility of a transition between the wild type and mutant state in a non-replicating cell can explain why in the actual phenomenon of adaptive mutations, the mutations do not take place all at once, immediately after the plating, but instead the number of mutant colonies steadily increases with time.

The remaining possibility of transition between Cw and Cm states makes the cell behavior in the environment $E_2$ somewhat analogous to the radioactive decay (Gamow, 1928; Gurney & Condon, 1928), as mentioned in the section 2. However, there are two important differences. First, there is no need to invoke the concept of tunneling for the description of this transition, as it can also be described as thermally activated barrier crossing[13]. Second, unlike in the simple α-decay case, the potential energy landscape can be modulated by changing the environments from $E_0$ to $E_1$ or $E_2$ or any other $E_N$ (see the Fig. 6 for the clarification of the original fluctuation trapping model, which also includes the case of the generic nonselective substrate (e.g. glucose) and acknowledges the generation of kinetic barriers between the $CB_E$ basis states after the change in environment).

**Figure 6**

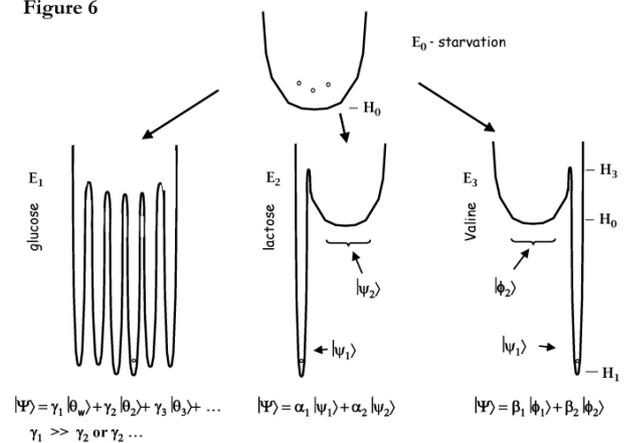

**7. The Q-genome approach and the *R-D-error* correlation. The difference from the Q-cell approach**.

The difficulty with analyzing the approach suggested by McFadden and Al-Khalili is that in the course of time it undergoes changes. As shown in section 4, their 1999 model employs the '*D-error* only' scenario and thus cannot be qualified as QMAM.

In the book (McFadden, 2000) and in their internet posting on arxiv.org (McFadden & Al-Khalili, 2001) the authors consider a possibility that, in addition to a *D-error*, an *R-error* may also be involved in the mechanism of adaptive mutation. They never state that both *R-error* and *D-error* have to be involved in the same scenario, let alone that there has to be a correlation between these errors. However, we will give the **Q-genome** model the benefit of the doubt, and assume that it can be somehow reformulated as involving an *R-D-error* correlation.

How is the *R-D-error* correlation possible in the framework of the **Q-genome** approach? The interpretation most consistent with the authors' giving high significance to the estimations of relaxation times for proton transitions *in vivo* (McFadden & Al-Khalili, 1999; McFadden & Al-Khalili, 2001) is that the nucleotide base under question has to stay in the same alternative tautomeric form for a sufficiently long time, so that it can be recognized by both RNA-polymerase and DNA-polymerase in the same erroneous way. Otherwise, the rapid proton transition back to its regular place would quickly change the rare tautomeric form of the base to its common form, making the probability of *D-error* independent from that of *R-error*. As a result, the memory of which event on the level of transcription has led to a useful change in the phenotype would be lost, and the cell will not have information on how to change its genome.

How, however, is it possible to keep the proton state from relaxing? The authors propose that the proton relaxation time should be on the order of 10-100 sec. However, this estimate was made on the basis of their original '*D-error* only' model. To have the *R-D-error* correlation, the proton relaxation time has to be much longer, as this scenario has to account for the whole reproduction cycle, which starts with the synthesis of the mutant RNA molecule (Figure 4 top right), includes the synthesis of the mutant protein (Figure 4 middle right) and ends by the eventual fixation of the mutation by DNA synthesis (Figure 4 bottom right).

Importantly, the **Q-cell** approach does not encounter this problem, as it does not require the proton to be 'frozen' in the rare tautomeric position for any significant period of time. In order to see that, the representation of the **U** state of the cell in the Cloning Basis can be further expanded using Schmidt decomposition as:

$$|\Psi\rangle = \alpha_w |C_w\rangle + \alpha_m |C_m\rangle =$$
$$= \alpha_w |P_w\rangle |R_w\rangle + \alpha_m |P_m\rangle |R_m\rangle \qquad (7.1)$$

, where $P_i$ describes the state of the proton (in regular $w$ and tautomeric $m$ positions, correspondingly), and $R_i$ describes the state of the rest of the cell. After tracing over the state of rest of the cell (R), the proton will be in the mixture of the states corresponding to normal and tautomeric positions, due to its entanglement with the rest of the cell:

$$\rho^P = \alpha^2_w |P_w\rangle\langle P_w| + \alpha^2_m |P_m\rangle\langle P_m| \qquad (7.2)$$

Thus, there are no strict limitations on the proton relaxation rates in the **Q-cell** approach. The proton position is allowed to change rapidly, as long as the rest of the cell changes with it. The 'Central Dogma' is cheated in a different way here – unlike in the **Q-genome** approach, it is not the proton position that preserves the memory about the *R-error*, but rather the *correlations* between the state of the proton and the state of the rest of the cell.

It is important to point out an additional difference between the **Q-cell** and **Q-genome** approaches. So far we used an implicit assumption that the source of the *R-* and *D-errors* was the tautomery of the nucleotide base located in the DNA template. This makes the **Q-cell** and **Q-genome** approaches look alike, since they both appear to require DNA to be in the state of superposition at some point. Invoking von Neuman's chain of observers (Von Neumann, 1955), one can then argue for an arbitrary character of the boundary between the 'unitary' and 'measurement' steps of the adaptation process (von Neuman II and von Neuman I classes, accordingly) and conclude that these approaches represent equivalent descriptions of the same scenario of measurement of the state of DNA. However, this conclusion would be wrong. The **Q-cell** approach is also compatible with a 'tautomery scenario' that does not involve superposition of DNA states at all. Consider a free precursor (e.g, ATP, dGTP etc) used by RNA- or DNA-polymerases to synthesize mRNA or daughter DNA molecules. Assuming that the nucleotide in DNA is in its regular form, the tautomery of a precursor molecule can equally lead to an **R-** or **D-error** via the same mis-pairing mechanism as discussed previously. In this scenario, the tautomery of the bases in the DNA template is not involved, hence there is no need to consider DNA in the state of superposition. This example demonstrates an important point. The **Q-cell** and **Q-genome** approaches cannot be the same idea in a different disguise – they are not only different in their formulation, but they are also not equivalent in their scope; in particular, the **Q-cell** can accommodate scenarios of adaptive mutations beyond the reach of the **Q-genome** approach.

## 8. Conclusion and outlook.

The present paper elaborates the conceptual framework for explanation of the phenomenon of adaptive mutations proposed previously by the author (Ogryzko, 1994; Ogryzko, 1997; Ogryzko, 2007; Ogryzko, 2008b). According to its main premise, the traditional molecular biological approaches will not suffice for the explanation of this phenomenon, and a better understanding of the physics of life will be needed. Moreover, one can expect that this better understanding might prove useful in revisiting some basic physical problems. In the past, biology contributed to many fundamental physical discoveries (such as energy conservation, electricity, etc). Nowadays, however, an input from the life sciences is not requested for tackling such fundamental physical problems as the problem of 'transition from quantum to classical' and the problem of thermodynamic irreversibility and the 'time arrow'. This might be too narrow a view, in light of the growing role of information concepts in the foundations of physics (Zurek, 1990). Living nature provides us with the first clear example of natural objects that utilize, store and process information for their own needs (Ogryzko, 1994), that is, the first example of IGUSes (Gell-Mann & Hartle, 1990; Saunders, 1993). It should not come as a surprise then that the future theoretical physics might draw as much inspiration from biology as it did 200 years ago, giving a new meaning to the famous words of David Hilbert '*Physics is too important to be left to the physicists*'. Below, some ramifications of the present paper will be pointed out and discussed from the above point of view.

### 8A. Quantum 'worlds/objects' vs. quantum 'properties'.

Life is traditionally presumed to belong to the realm of classical physics. Accordingly, it is generally believed that the transition from quantum to classical descriptions has to be dealt with before theoretical investigations of life can start. This view is based on the fact that living organisms are macroscopic and warm objects. The unfortunate misconception that the larger an object is, the more classically it behaves has partially historical roots. The hydrogen atom, due to its simplicity, was the first physical object where the inadequacy of the classical explanations could be unambiguously demonstrated. However, there is nothing in the conceptual apparatus of quantum theory, nor in our experience, that precludes the idea that some properties of macroscopic objects could be non-classical, i.e., described with the use of noncommuting operators.

Of all current approaches to the transition from 'quantum to classical', the environment induced decoherence (**EID**) approach of Zeh, Zurek and their collaborators (Zeh, 1970; Zurek, 2003; Zurek et al., 1993) serves best to illustrate the above point. The state selection via decoherence (einselection) depends critically on the interaction of the system with its environment. Thus, it is not the size of the system but rather the mode and the symmetries of its coupling with environment that will determine which of its states will survive. Then, no matter how large the object is, noncommutativity will reveal itself whenever a change in environment causes a change in the spectrum of preferred states (section 6A). According to this argument, and given that Life is known to have very intricate connection to its inhomogeneous and changing environment, the phenomenon of biological adaptation appears to be fitting very naturally into the general **EID** scheme (Ogryzko, 2008a; Ogryzko, 2009). The novelty of our approach is in making an explicit use of the dependence of the spectrum of preferred states from the structure of the environment of the object studied, implicit in the einselection idea.

Overall, I feel that the problem of 'quantum' versus 'classical' might greatly benefit if reformulated in a new way: instead of the opposition between classical and quantum *worlds* (inhabited by classical or quantum *objects*, correspondingly), we should consider classical and quantum *properties* of the same objects. Even the electron, in addition to its many quantum properties, has classical properties as well – such as its charge, and intriguingly enough, the environmentally induced decoherence has been implicated in its emergence (Giulini, 2000 ;

Giulini et al., 1995). Conversely, given that quantum theoretical formalism is considered a fundamental language for description of physical reality, many objects that apparently have been a subject of the 'quantum to classical' transition can have some of their remaining properties described by noncommuting operators. This point is essential for understanding the meaning of the **Q-cell** hypothesis. Obviously, many properties of the bacterial cell are classical – the position and momentum of its center of mass in physical space is case in point. Nevertheless, the most interesting things about the cells are the processes taking place at the molecular level. The description of the intracellular dynamics from first principles has to start with quantum mechanics, treating a state of the cell as a density matrix operating on the high-dimensional Hilbert space that specifies positions of every nucleus and electron in it. There are many reasons to expect that the non-classical features of the quantum mechanical description (entanglement, for example) will be found to remain relevant even after environment is taken into account and all other possible approximations and simplifications are carried out (Ogryzko, 2008a; Ogryzko, 2009).

**8B. Euclidean approach to the reproduction problem.**

Section 6B discussed why the central to the Q-cell approach idea of 'cell in a state of superposition of mutant and wild type states' should not be confused with the Schroedinger cat case, and how this idea can be described with the operator language. What could be the mathematical form of the operator $O_{E1}$, which represents the ability of a cell to grow on a specific substrate (for the case of lactose, we used notation *Lac* for such an operator (Ogryzko, 2007; Ogryzko, 2008b)) and is used in the description of the cell in the superposed state? The discussion below (and in the section 6C) indicates that it should be related to the Hamiltonian operator.

It has long been recognized that the connection between the time and space derivatives of the state vector, expressed by the Schroedinger equation, is formally equivalent to the description of *heat redistribution*, but occurring in imaginary time (Fenyes, 1952; Nelson, 1966). This paper suggests that an alternative interpretation of the same equation is possible, namely as '*reproduction* in imaginary time' (Section 6C). Regardless of its application in the present work, the proposed idea can find other potential uses in the field of theoretical biology and the theory of self-reproducing automata. One such application could be a novel strategy for the theoretical description of reproduction process at the molecular level. I term this strategy 'Euclidean approach' (Ogryzko, 2009), for the reasons outlined below.

The mathematical description of reproduction as a physical process is notoriously difficult. Among the conceptual roadblocks are the open character of the reproducing object as a physical system and the generally irreversible nature of the reproduction process. Even more intimidating is the problem of description of the shift from a single mother object to two daughter objects. When describing dynamics of the reproducing system on the molecular level as a movement of a point (or a finite volume) in a corresponding high-dimensional state space, how to represent this dynamics in a way that would describe the transition from a single object to two objects of a similar type?

Admittedly, physicists and mathematicians feel far more comfortable with closed or isolated systems, described by deterministic reversible equations (mathematically, groups are much better understood and tamer objects compared to semigroups). From this perspective, the 'reproduction in imaginary time' is much closer to home when compared to the 'real time' reproduction, offering all the technical advantages of a closed system evolving in a reversible and deterministic way, and also needing to keep a track of a single object only.

Accordingly, I propose to split the task of the physico-mathematical description of cell reproduction at the molecular level into two steps. We start with considering the cell undergoing reproduction in imaginary time. As argued above, the search for consistent solutions of the equations describing this state is expected to be technically simpler in this case, and will be essentially reduced to finding the eigenstates of the Hamiltonian describing the dynamics of the **U** state (Table 1). At the second step, Wick rotation can be performed, which can be expected to yield the description of real reproduction by continuing the discovered solutions analytically to the 'real time' coordinate. As a motivation of the proposed strategy, one can consider its direct analogy to Euclidean methods in quantum field theory (QFT) and quantum gravity (QG) (Hawking, 1988). In this approach, a substitution of *t* by *it* often yields a problem in real Euclidean coordinates, which is easier to solve, and can be used for a search of the

'real time' solutions after reverting the Wick rotation.

Importantly, unlike in QFT and QG, where Euclidean approach appears as a formal mathematical trick, in the case of cell reproduction both Wick rotation and 'imaginary time' process have their counterparts in the real world. In my proposal (section 6C), reproduction in imaginary time describes intracellular dynamics in a starving cell, whereas Wick rotation corresponds to switch from the 'starving' to the 'proliferation' regimes, which can be achieved by addition of a substrate to the cells. Accordingly, compared to QFT and QG, my proposal has an advantage of being more amenable to experimental verification. Given that 'imaginary time' allows reproduction of superposed **CB$_E$** states (and moreover, it is these superpositions that are einselected (as a **U** state) due to the continuous variations at the genomic level such as base tautomery (see 6C)), one promising research avenue to validate the Euclidean approach could be the study of the phenomenon of adaptive mutations.

It is beyond the scope of the present paper to produce a full account of the Euclidean approach for physico-mathematical description of cell reproduction at the molecular level. I will only comment on one aspect of this program – the need to find an alternative ontological justification for the use of the Laplacian operator $\nabla^2$. This operator plays a central role in many fields of mathematical physics. The question of why it is used for the description of unitary dynamics could be seen in the context of a more general problem of the nature and the origin of the physical laws. Commonly, a stochastic process (random walk, diffusion) is presumed to underlie the connection between the time and space derivatives of the state vector, described by $\nabla^2$. Confirming the intuition of the ancient atomists, this suggests that, on a fundamental level, randomness underlies quantum-mechanical description of the physical world and is at the core of other laws of physics (Nelson, 1966). But is there another fundamental property or process that could serve as an alternative to the 'stochasticity' property in the justification of Laplacian? In the context of the role of EID in biological adaptation, I will limit myself to one idea and pose a question of what operations can be performed on the description of a physical system in order to take into account its relationship to its environment. One can consider two such operations (Table 2). The first one is coarse graining, usually justified by the impossibility for an external observer to know everything about the state of the system. Due to the information loss incurred by coarse graining, this procedure naturally introduces a stochastic element into the dynamics of a closed system. This is a way to arrive at the standard 'diffusional' interpretation of the Laplacian operator $\nabla^2$. Importantly, however, the desired formal expression in imaginary time **(6C.1)** can be similarly obtained via a Wick rotation[14] if we start from a slightly different expression:

$$(\partial/\partial t)\Psi = - D\nabla^2\Psi , \qquad D > 0 \qquad (8.B1)$$

describing an *amplification of local differences* in occupation between neighboring states, instead of their diffusional smoothing out with time. How is it possible to justify this expression **(8.B1)**, i.e., what operation on the description of a system could naturally lead to such a 'sharpening' process? One can expect that, as opposed to the coarse graining procedure, it should be an operation which leads to an increase in certainty about the system, instead of the loss of information about it. This increase in certainty can be achieved by including our system within a larger system, and thereby adding to its description new degrees of freedom, previously unaccounted for. Notably, this alternative way to take environment into account is particularly appealing when dealing with biological systems, due to the known relativity of the physical boundary between biological object and its surroundings, manifested in exchange of matter and energy between them. It remains to be explored what connection this idea has with our suggestion (6.C) that the addition of external substrate to the starving cell leads to the switch from the regime of 'imaginary time reproduction' to the one of 'real time reproduction'.

## 8C. 'Cheating' the Central Dogma of Molecular Biology.

The notion of a 'starving bacterial cell', one of the main subjects of this paper, presents two conceptual challenges for Molecular Biology. First, starving cells appear to be able to adapt to their environment in a Lamarckian fashion, thus putting the Central Dogma of Molecular Biology in doubt. The second challenge is the apparent ability of these cells to survive for several days without nutrients. In terms of physics, how one can explain this property in the light of the commonly accepted view of biological systems as being in a far from equilibrium state and thus requiring constant energy expenditure for the maintenance of their ordered structure? The implications of this paper is that these two challenges are related, and that both problems can be addressed by proposing that intracellular processes in a starving cell can be approximated by unitary, i.e., physically conservative, dynamics.

The two problems are related, insofar as the unidirectional character of causal influences in the cell requires that intracellular processes are irreversible. However, when described by unitary dynamics, the intracellular dynamics becomes *reversible* and *conservative*. Consequently, the stability (i.e., survival) problem can be addressed along the lines of Schroedinger's suggestion that, physically, the operation of a living organism resembles the operation of a mechanical system, being 'largely withdrawn from the disorder of the heat motion' (Blumenfeld, 1981; Schroedinger, 1944). As far as the 'unidirectional information flow' is concerned, the fact that all information is preserved in a unitary process entails that the starving cell should be able to keep track of what event at the genotypic level (such as ***R-error***) has led to the appearance of a useful change at the level of phenotype (Ogryzko, 1997). This indicates the crucial point where the 'Central Dogma of Molecular Biology', forbidding the information flow from phenotype to genotype, loses its adequacy.

The language of quantum information theory can help to illustrate the limitations of the common notions of causality and control when they are considered in the context of unitary dynamics (Janzing & Decker, 2007). The direction of information flow becomes clearly dependent on the representation basis in the case of unitary dynamics, but on the other hand, the basis itself can be arbitrarily chosen. An elementary example of this relativity is the symmetry of the controlled-not (CNOT) gate (Nielsen & Chuang, 2000; Zurek, 2003) :

$|0\rangle_A|0\rangle_B \rightarrow |0\rangle_A|0\rangle_B$
$|0\rangle_A|1\rangle_B \rightarrow |0\rangle_A|1\rangle_B$
$|1\rangle_A|0\rangle_B \rightarrow |1\rangle_A|1\rangle_B$
$|1\rangle_A|1\rangle_B \rightarrow |1\rangle_A|0\rangle_B$ (8.1)

This unitary gate has a qubit A as control and qubit B as target. A CNOT gate allows us to transmit one bit of information from A to B: to do this, one initializes B to the basic state $|0\rangle$ and chooses one of the states $|0\rangle$ or $|1\rangle$ for the system A. After the action of the CNOT gate on the joint system, we obtain B in $|0\rangle$ or $|1\rangle$ depending on which state we have chosen for A. However, we can also choose another basis for the description of the same system (the so-called Hadamard basis):

$|+\rangle \equiv (|0\rangle + |1\rangle)/\sqrt{2}, \quad |-\rangle \equiv (|0\rangle - |1\rangle)/\sqrt{2}$ (8.2)

In this basis:

$|+\rangle_A|+\rangle_B \rightarrow |+\rangle_A|+\rangle_B$
$|-\rangle_A|+\rangle_B \rightarrow |-\rangle_A|+\rangle_B$
$|+\rangle_A|-\rangle_B \rightarrow |-\rangle_A|-\rangle_B$
$|-\rangle_A|-\rangle_B \rightarrow |+\rangle_A|-\rangle_B$ (8.3)

therefore, after we have initialized A to the state $|+\rangle$, the same gate will allow us to transmit one bit of information from B to A, i.e., the control and the target parts of the gate have interchanged roles.

This example suggests that, if intracellular dynamics can be described as a unitary process, the cell would not need any special mechanisms in order to cheat the Central Dogma of Molecular Biology. The same molecular hardware (transcription, translation and replication apparatus) will be sufficient in order to provide the Lamarckian feedback 'from phenotype to genotype'. Usually, it is DNA that is considered as the control part of the cell 'A', switching between alternative states ($|wild\ type\rangle$ and $|mutant\rangle$), and thus determining the state of the rest of the cell, which plays the role of the target part 'B'. But in the case of unitary dynamics, a different basis is equally legitimate. In this alternative 'Hadamard-like' basis the state of DNA (either symmetric or asymmetric superposition of the $|wild\ type\rangle$ and $|mutant\rangle$ states) will be on the receiving end of the information flow, i.e., phenotype will be controlling the genotype. Importantly, the ambiguity in the basis choice is lost, and the symmetry (bi-directional character) of the information flow is broken in regular growth conditions, i.e., when the intracellular processes are irreversible. It is in the starving cells that the unexpected subtlety in the connection between genotype and phenotype are best revealed, consistent with the fact that this experimental model is proving to be most fruitful for the study of adaptive mutations.

### 8D. Q-cell and Q-genome approaches

The standard formalism of quantum theory distinguishes between two classes of physical processes (Von Neumann, 1955). The so called von Neuman II processes are deterministic, reversible, conservative and are described by unitary equations. The other class of processes corresponds to a measurement (von Neuman I process) and is related to the much debated issues of physical irreversibility, the 'quantum to classical' transition, and the role of the observer. The connection between the two is still poorly understood and, in fact, remains a fundamental problem of contemporary physics. One can argue, however, that quantum theory owes its success largely to the art of recognizing which part of the phenomena can be comfortably described as the IInd class, and which part will carry the burden of the interpretational/foundational problems and has to be assigned to the Ist class. From this perspective, the main challenge for a **QMAM** is how to capitalize on this distinction in approaching the problem of adaptive mutations, i.e., how to describe this phenomenon in terms of the IInd and Ist classes of processes.

This paper compares two such attempts, termed here **Q-cell** and **Q-genome** (Table 3). The logic of the **Q-cell** approach starts with analysis of operational limitations on what can be observed considering an individual biological object (e.g. a bacterial cell). It then proposes to apply the formalism of quantum measurement for the description of bacteria plating experiments, with a von Neuman II process describing the state of the starving cell, and the bacteria plating procedure corresponding to a von Neuman I process. As a result, it arrives at the scenario of *'selection among virtual states of the individual organism'*, a novel adaptation scheme characterized by the impossibility of separating the variation and selection steps of the adaptation process (i.e., the dependence of sampling space from the conditions of observation) (Ogryzko, 1997; Ogryzko, 2007). To the contrary, the approach of McFadden and Khalili, referred to here as **Q-genome**, focuses on the state of DNA and considers the cell as a device measuring the state of its DNA. Regrettably, McFadden and Al-Khalili do not recognize the essential difference between the two approaches and mischaracterize my model as a **Q-genome** approach (McFadden & Al-Khalili, 1999).

This paper argues that the **Q-cell** approach has several advantages over the **Q-genome** approach in accounting for the phenomenon of adaptive mutations (Table 3). The fluctuation trapping process is more adequately described in the language of the **Q-cell** approach, inasmuch as the description of fluctuations (virtual mutations, in the previous terminology (Ogryzko, 2007)) cannot be reduced to the description of the variations in the state of DNA only, but also have to include certain compensatory changes in the rest of the cell. Accordingly, the focus of the **Q-genome** approach on a part of the cell, instead of the dynamics of the whole cell, is prone to lead in unproductive directions. This is testified by its authors' giving high significance to the relaxation times of proton position (a misleading and irrelevant point, as argued in the section 7), and also by the fact that, despite their claim, the first model of McFadden and Al-Khalili does not qualify as a QMAM (sections 3 and 4).

As may be seen from section 7, the **Q-genome** approach is difficult to reconcile with the general einselection-based scheme of adaptation process (section 6A). Whereas environmental decoherence plays positive role at both stages of adaptation in the framework of the **Q-cell** approach (before and after the change of environment $E_0 \rightarrow E_2$), it remains a problem for the **Q-genome** approach (at the 'before' stage). The advantage of the **Q-cell** approach is due to the fact that **EID** makes it easier to preserve the coherence of the state of a whole cell compared to the coherence of the state of its genome alone. This seemingly paradoxical situation arises owing to the fact that it is not the size of the system that matters for **EID**, but rather the strength and mode of its coupling with the environment. The interactions of DNA with other parts of the cell, such as the replication and transcription apparatus, are essential for its role in the cell and cannot be neglected in any description of its functioning; and these interactions will destroy a superposition of DNA states in the blink of an eye (i.e. convert these states into the 'improper mixtures'). On the other hand, the interactions of a starving cell with its environment are weaker and less essential. Hence the preservation of coherence via einselection is more realistic in the case of a cell than in the case of a DNA molecule inside it.

Finally, as shown above (section 7, end), the **Q-cell** approach can accommodate scenarios of adaptive mutations that cannot be considered by the **Q-genome** approach. The broader applicability of the **Q-cell** approach to biological adaptation compared to the **Q-genome** can be further illustrated on the following example. In addition to '*hard adaptation*', which involves changes at genetic levels, one can also consider '*soft adaptation*', which takes place at the epigenetic level only. Epigenetic adaptation can be considered alone and outside of the evolutionary context, simply as a manifestation of the plasticity of an individual organism in response to its environment. However, it can also play a role in the evolutionary adaptation scheme, associated with the names of C. Waddington and J. Baldwin (Baldwin, 1896; Waddington, 1953). This scheme acknowledges innate plasticity of individual organism and considers adjustment at the epigenetic level as the first step in the process of adaptive evolution. Only at the second step these epigenetic changes are fixed at the genetic level. Soft adaptation does not require proliferation of an organism, and, by definition, does not involve genomic variations at any stage. However, it can be described equally well by the einselection adaptation scheme (Section 6A) (Ogryzko, 1994).

**8E. Lamarck or Darwin?**

The evolutionary theory of Lamarck (Lamarck, 1809), historically the first explanation of biological adaptation and diversity based strictly on natural laws, was supplanted by Darwinism in the biology of the XXth century. The Lamarckian principle of the inheritance of acquired characteristics had fallen into disrepute due to the lack of empirical evidence for molecular mechanisms that would implement direct feedback from phenotype to genotype at the level of an individual organism. However, both Darwin and Lamarck operated with classical concepts. The progress in physics of the XXth century, and of quantum mechanics in particular, enriches our understanding of the concepts of causality and control. It gives more credibility to the Lamarckian notions, by suggesting the existence of more subtle links between genotype and phenotype than could be expected from the classical view-point.

Is the vote recount long overdue in the century-old dispute between Lamarckism and Darwinism? Arguably, the very difference between the two paradigms appears to be blurred in the proposed approach. On the face of it, the idea of 'selection of virtual mutations' looks very much like a Darwinian concept. Should we call a truce then and submit that both Lamarck and Darwin could have been right? Although tempting, this would not be the best way to proceed. Science benefits most from keeping clear demarcation lines between different paradigms (*and redefining them, if necessary*). This practice keeps the scientific discourse going, in part by stimulating development of predictions and experimental tests to distinguish between various alternatives. Population-level thinking and the separation between the variation and selection steps of biological evolution have been the cornerstones of Darwinism from its conception. Assimilating the idea of 'selection of virtual mutants' would devalue Darwinism, completely depriving it of its predictive power. Being in favor of establishing clear demarcation lines, I propose to define the essential difference between Darwinian and Lamarckian paradigms of biological evolution as *what is taken as the true object of evolutionary dynamics*. Darwinism considers a population (of genes or organisms) as the only real object of evolutionary dynamics, reducing the elements of population (individual organisms or genomes/genes) to rigid and disposable units, good only for being discarded or kept for the next generations. Lamarckism, on the other hand, admits that an individual biological object has enough plasticity and resources to adapt and thus to contribute directly to evolutionary adaptation, thus leaving less need for the notorious Darwinian competition and the struggle for survival. I hope that drawing this clear distinction will give a fresh impetus to evolutionary studies and help to establish new productive directions for experimental and theoretical research.

**Acknowledgments**:

The author thanks Murat Saparbaev for discussion, Bernard d'Espagnat for comments and discussion, Henry Stapp and Roland Omnés for comments, Linda Pritchard for the proofreading and comments, Marc Lipinski for the support and encouragement.

**Abbreviations list**

**QMAM** – quantum model of adaptive mutation**,**
**EID** – environmentally induced decoherence,
**MFAK99** – model of McFadden & Al-Khalili, 1999 (McFadden & Al-Khalili, 1999),
**MBB** – molecular biology basis**,**
**PRB** – phase rotation basis**,**
**CB** – cloning basis**,**
**QFT** – quantum field theory**,**
**QG** – quantum gravity.

# Endnotes

[1] Moreover, this scenario was rejected by Cairns as a 'leaking mutant model'

[2] It might be convenient, when describing the cell in the starving conditions, to distinguish between the notions of 'einselected state' and a 'preferred state'. The notion of 'einselected state' (**U** state) would refer to a state selected as a result of interaction with the environment - and given the uncertainty caused by the coupling with the environment, the **U** state has to be a mixture of preferred states (i.e., the elements of the diagonalized density matrix describing the einselected state **U**). Thus, we reserve the notion of 'preferred state' for any one of the pure states contributing to the einselected state **U**.

[3] In spite of the fact that the einselected state **U** is a mixture of preferred states, each of the preferred states, considered individually, can be represented as a superposition of the wild type and mutant. Therefore we can safely state that the einselected **U** state is in the state of superposition, since it does not matter in which of the preferred states the cell actually is. The language of the off-diagonal terms of a density matrix is convenient to describe this situation. We simply have to state that the density matrix describing the original einselected state **U**, is not diagonal in the basis of the wild state W and mutant state M, i.e. will contain off-diagonal terms, describing interference between the W and M states (more about it later in 6D).

[4] It is beyond the scope of this paper to describe the mathematical form of this operator, but the discussion in the sections 6C and 8B suggests that it should be related to the Hamiltonian operator.

[5] The system does not have to be in a pure state in order to undergo unitary evolution. As an example one can consider 'decoherence free subspaces' (DFS), discussed in the quantum computation theory as a way to protect quantum information processing from EID (Zanardi & Rossetti, 1997). Intriguingly, there is a strong mathematical relation between the DFS and the formalism of preferred states, as discussed in (Zurek, 2003)

[6] Description of a bound state, such as an electron atomic orbit, also includes a potential V(x). However, given that the potential energy terms can often be eliminated by a coordinate (canonical) transformation, the consideration of the simplest version is sufficient for our discussion.

[7] And, given that in the quantum field theory presence of virtual particles makes every system appear composite, this consideration has a general relevance.

[8] The use of complex numbers is crucial for understanding why the intracellular dynamics in a starving cell (enzymatic and diffusion-driven transitions between the **MBB** states) does not eventually lead to degradation of its ordered state. Such degradation would be inevitable if we were limited to real numbers only – a reaction-diffusion system cannot maintain its order if there is no flow of energy through it. However, the use of complex numbers alleviates the problem – according to the Fundamental Theorem of Algebra, a stationary solution (a stable **PRB** state) will always exist if complex numbers are allowed.

[9] , which also could include continuous Wick rotations, spanning complex values of the time variable.

[10] Some epigenetic information could be amplified as well and thus would introduce a more fine structure into the C basis, but it is not important for the current discussion.

[11] For simplicity, we do not consider lethal mutations.

[12] More on the role of environmentally induced decoherence in generating superselection rules one can read in (Giulini, 2000 ; Giulini et al., 1995)

[13] In other words, quantum mechanics is working its magic in a different place here – we do not require it to describe the transition between the wild type and mutant states, but we absolutely need it for explaining how only these states are selected in the new conditions.

[14] This Wick rotation will differ from the original one by the sign in front of '*i*'

**Figure Legends:**

**Figure 1. Fluctuation trapping model I, and its relation to measurement.** A simplified version of the fluctuation trapping model, that captures the essence of its relation with the measurement procedure. A particle is delocalized over an even potential surface (Top). In order to observe if it is located in a particular place (M1 or M2), we generate a deep potential well in this location (Middle). With some probability the particle will fall into the well (Bottom). When it does fall, it loses energy ($\Delta H$), which can be detected by an observer A as a photon emission ($\hbar \nu = \Delta H$). This is an irreversible process, by which the particle is trapped in the position where we wanted to observe it. Choice of a potential well in a different place will lead to the particle eventually being trapped in a different location. By choosing to generate a potential well in particular place, the corresponding sampling space is generated by breaking the set of all potential positions of the particle into two classes (M and W), that correspond to two different outcomes of the observation. Applying this general model to adaptive mutagenesis, the creation of a potential well corresponds to addition of a specific substrate (lactose or other nutrient) to the plate with bacteria, and the trapping of the particle – to an appearance of a colony on the Petri dish. A different location of the well will correspond to a different sampling space – a different way to break the set of all positions into two classes ($\phi_1$ and $\phi_2$ instead of $\psi_1$ and $\psi_2$), consistent with the main feature of this model of adaptation – inability to separate the variation step from the selection step.

**Figure 2. *R-error* only model.** Cytosine (and other nucleotides) have two tautomeric forms, due to proton transition from the 4-amino N to the 3-imino N, accompanied by reconfiguration of the electron structure from an endocyclic to an exocyclic double bond structure. Accordingly, the state of the system containing a nucleotide base is written as a superposition of the states corresponding to the regular and tautomeric forms of the base. Recognition of the tautomeric form of cytosine (right) by RNA-polymerase will lead to generation of mRNA with A in place of G, and its subsequent translation will generate an arginine (R) to histidine (H) substitution in the aminoacid sequence of the encoded protein. This model requires an additional step of fixation of the useful change in DNA sequence.

**Figure 3. *D-error* only model.** Recognition of the tautomeric form of cytosine (right) by DNA-polymerase will lead to generation of a mutant DNA copy. To test its usefulness at the phenotypic level requires faithful transcription and translation of the mutant DNA copy, leading to generation of a protein with a useful substitution in aminoacid sequence.

**Figure 4. The *R-D-error* correlation.** Recognition of the tautomeric form of cytosine (right middle) by RNA-polymerase (***R-error***) will lead to generation of mutant mRNA and a mutant protein. Recognition of the tautomeric form of the same cytosine by DNA-polymerase (right bottom, ***D-error***) will lead to generation of a mutant DNA copy and fixation of the useful ***R-error***.

**Figure 5.** We apply forces $\mathcal{PF}1$ and $\mathcal{PF}2$ to pull apart the crystal lattice **L** until it breaks in two pieces **L'** and **L'**. We can see application of the forces it as a change from environment **E₀**, where the lattice was stable, to environment **E₁**, where it becomes unstable. Language of superposition helps to describe the choice of the exact location of the breaking point (labeled by * is a sample of these points). Breaking point is expected to be the place where the bonds holding the crystal together are most distorted, i.e. have the highest energy. In the phonon description, this point would correspond to the phonon position. As discussed in the text, in the environment **E₀** the lattice is in the state of superposition of the eigenstates of the phonon position operator. The environment **E₁** can distinguish between these alternative states of the lattice, thus the above superpositions are expected to be quickly destroyed in **E₁**, which corresponds to breaking the lattice up.

**Figure 6. Fluctuation trapping model II:** Clarification of the fluctuation trapping model, described in a simplified form in the Figure 1. It includes also the case of a generic nonselective substrate (glucose, represented by environment E₁) and acknowledges the presence of kinetic barriers (of a finite energy H₃) separating different states after the change of environment. Top – the starving cell in **U** state. The potential landscape is not flat as in Figure 1, reflecting the notion that, physically, the einselected state of the cell corresponds to a bound state protected by kinetic barriers (Ogryzko, 2009). Bottom left – after an addition of a nonselective substrate (glucose), many potential wells are generated, each corresponding to a different genetic sequence and a different element of the **CB$_{E1}$** basis: $|\theta_w\rangle, |\theta_2\rangle, |\theta_3\rangle \ldots$ , with $|\theta_w\rangle$ designating the wild type genome. Given that the probability of a spontaneous mutation per genome is less than $10^{-8}$ and the size of E.coli genome is $4.6 \times 10^6$ base pairs, the contribution $\gamma_1$ of the $|\theta_w\rangle$ will be much larger than other $\gamma_i$. Bottom middle – addition of lactose will generate one potential well, corresponding to a mutant able to grow on lactose $|\psi_1\rangle$ (for simplicity, we assume that only one genetic sequence will be able to give growth in these conditions). For the rest of the sequences (the wild type included), it will not be possible to distinguish between them without destroying the cell. Accordingly, as long as the cell is alive and is in the environment **E$_2$**, all of the remaining sequences will have to be combined into one basis state $|\psi_2\rangle$. For a cell in this state, a finite probability to transit to the $|\psi_1\rangle$ will remain, leading to a steady increase of mutant colonies on the lactose plate with time. Bottom right – similarly, any other selective condition (in this case, valine instead of lactose) will correspond to generation of a different potential landscape and a different set of basic states ($|\phi_1\rangle, |\phi_2\rangle$).

**Tables:**

**Table 1. Comparison between reproduction 'in real time' and 'in imaginary time'**

|  | **Real Time** | **Imaginary Time** |
|---|---|---|
| Physics | Irreversible | Reversible |
|  | Dissipative | Conservative |
| Mathematics | Nonlinear, described by a semigroup | Linear, described by a group |
| Basis for description | Preferred basis exists (CB) | Preferred basis ambiguous[a] |
| Information flow | Unidirectional, from genome to the rest of the cell | Bidirectional, depends on the choice of basis |
| Cloning | Only cloning basis (CB) | Arbitrary states can be cloned[a] |

[a]: See Appendix

**Table 2. Comparison between 'diffusion' and 'reproduction' justification of Laplacian**

|  | **Diffusion** | **Reproduction** |
|---|---|---|
| Behavior in time | Negative exponent | Positive exponent |
| Dynamic in space of states | Differences decay (blur) with time | Differences amplify (sharpen) with time |
| Information about the state | Loss of information | Gain of information |
| Operation on the description of the system | Coarse graining | Inclusion within a larger system |
| Quantum Information procedure (density matrix procedure)[a] | Tracing over | Purification |
| Category theory description[a] | Surjection/epimorphism/factor structure | Injection/monomorphism |

[a]: Last two lines are not discussed in the text, but include additional considerations on the mathematical aspects of the differences between the two justifications of the Laplacian operator.

**Table 3. Comparison of the Q-genome and Q-cell approaches.**

|  | Q-genome | Q-cell |
|---|---|---|
| Justification | Proton tunneling leading to base tautomery | Observational limitations at the level of an individual cell |
| Measured object | DNA | Cell |
| Measurement device | Cell | Environment/Petri dish with agar and substrate |
| Explanatory principle | Inverse Zeno effect | No separation between variation and selection steps - the spectrum of variations (sampling space) depends on environment |
| Role of decoherence | At the selection step only | Before selection – stabilization of the W and M superposition, during selection – destabilization of W and M superposition |
| Involvement of base tautomery | On DNA level only | Both DNA and the precursor could be involved |
| Epigenetic/Soft adaptation | Cannot be described | Can be described |

**Appendix**

We consider here an alternative way to harmonize the idea of 'reproduction in imaginary time' with the non-cloning theorem. One can argue that since the starving cell in the U state does not undergo actual self-reproduction and no external substrate is consumed, the arguments that forbid the copying of arbitrary quantum states simply do not apply to the case of 'reproduction in imaginary time'.

First we present a recapitulation of the proof of the non-cloning theorem, illustrating the crucial role played by the external substrate in the 'real time replication' scenario:

Assume that we can clone a state $\psi$ of a system A ($|\psi\rangle_A$), by converting, via a unitary process H, the blank state $|e\rangle_B$ of another system B to the identical state $|\psi\rangle_B$:

$$H|\psi\rangle_A|e\rangle_B = |\psi\rangle_A|\psi\rangle_B \quad (A.1)$$

Assume that another arbitrary state $\phi$ can also be cloned via the same H:

$$H|\phi\rangle_A|e\rangle_B = |\phi\rangle_A|\phi\rangle_B \quad (A.2)$$

By definition, the unitary operator *H* preserves the inner product:

$$\langle e|_B\langle\phi|_A|\psi\rangle_A|e\rangle_B = \langle e|_B\langle\phi|_A H^*H|\psi\rangle_A|e\rangle_B =$$
$$= \langle\phi|_B\langle\phi|_A|\psi\rangle_A|\psi\rangle_B$$

so that

$$\langle\phi|\psi\rangle = \langle\phi|\psi\rangle^2 \quad (A.3)$$

which is in general not true. Thus, no unitary operation can clone arbitrary states.

Now let's proceed with our argument. An important part of the above proof is that the substrate B is taken to exist *independently* from the system A. In this case, the initial state of B can always be represented by the same blank $|e\rangle_B$, regardless of the state of A. In other words, if we consider a linear combination of two different states of a composite system (A+B): $|\psi\rangle_A|e\rangle_B$ and $|\phi\rangle_A|e\rangle_B$, the blank state can be 'taken out of brackets', and the state of the composite system (A+B) will appear as a product state:

$$|\psi\rangle_A|e\rangle_B + |\phi\rangle_A|e\rangle_B = (|\psi\rangle_A + |\phi\rangle_A)|e\rangle_B \quad (A.4)$$

Consider now what will happen with the proof if we relax the requirement of independence of B from A and allow that for every state of A $|\psi\rangle_A$ there exists a dedicated 'blank' state of B $|e^\psi\rangle_B$, such that:

$$H|\psi\rangle_A|e^\psi\rangle_B = |\psi\rangle_A|\psi\rangle_B \quad (A.5)$$

For any other arbitrary state $\phi$ there will be another 'blank' $|e^\phi\rangle_B$, and the same H will give:

$$H|\phi\rangle_A|e^\phi\rangle_B = |\phi\rangle_A|\phi\rangle_B \quad (A.6)$$

Taking again the inner product:

$$\langle e^\phi|_B\langle\phi|_A|\psi\rangle_A|e^\psi\rangle = \langle e^\phi|_B\langle\phi|_A H^*H|\psi\rangle_A|e^\psi\rangle_B = \langle\phi|_B\langle\phi|_A|\psi\rangle_A|\psi\rangle_B$$

we obtain the condition of clonability:

$$\langle\phi|\psi\rangle = \langle e^\phi|e^\psi\rangle \quad (A.7)$$

Thus, the cloning of arbitrary states might not be forbidden by a unitary operation H, if the state of the 'substrate' B is always pre-correlated with the state of the cloned system A. How realistic is it to demand such a dependence between systems A and B? Formally, this requires that the state of the composite system (A + B) cannot be represented as a product state, but only as $(|\psi\rangle_A|e^\psi\rangle_B + |\phi\rangle_A|e^\phi\rangle_B)$. Thus we will have to require that these systems are entangled. But then B cannot be considered as an 'external substrate' with regard to A, being not separable from it (i.e., its state cannot be 'taken out of brackets' in the description of the state of the composite system 'A+B' at the beginning of the cloning process). Choosing this route to harmonize the idea of 'reproduction in imaginary time' with the non-cloning theorem, one can consider an alternative formulation of the non-cloning theorem, which would also accommodate the case of imaginary-time reproduction:

*Cloning of arbitrary quantum states is allowed in 'imaginary time' but becomes restricted to cloning of only orthogonal states after Wick rotation, i.e., in 'real time'.* **(A.8)**

The two alternative ways to harmonize the idea of 'reproduction in imaginary time' with the non-cloning theorem both could be valid and reflect a difference in what is considered as a reproducing entity, i.e., how the environment is taken into account in our description. It is tempting to relate this issue to the known controversy surrounding the possibility of coherent superpositions of photon number eigenstates in a radiation field. We refer the reader to the review (Bartlett et al., 2007), which offers an insight into this problem by taking into account the role that reference frames play in the description of quantum systems, i.e. whether they are considered from the point of view, which is 'internal' or 'external' to the system.